\documentclass[aps,prl,twocolumn,superscriptaddress,nofootinbib]{revtex4-1}
\usepackage{amsmath}
\usepackage{graphicx}
\usepackage{amsbsy}
\usepackage{amssymb}
\usepackage{palatino,avant,graphicx,color}
\usepackage[none]{hyphenat}
\usepackage{tikz}
\usepackage{xy}
\usepackage{comment}
\usepackage[none]{changes}
\usepackage{wrapfig}
\usepackage{float}
\usepackage{tabularx}
\usepackage{natbib}
\usepackage{hyperref}
\usepackage{mathtools}
\usepackage{palatino} %a package needed to introduce non-italic \mu (see also https://tex.stackexchange.com/questions/144569/non-italic-math-mode-symbols)
\begin{document}

\title{Subnanometer imaging and controlled dynamical patterning of thermocapillary driven deformation of thin liquid films}

 \author{Shimon Rubin}
 \thanks{These two authors contributed equally}
 \email{rubin.shim@gmail.com}
% \email{bjhong@eng.ucsd.edu}
 \author{Brandon Hong}
 %\thanks{These two authors contributed equally}
% \email{rubin.shim@gmail.com}
 \author{Yeshaiahu Fainman}
 \affiliation{Department of Electrical and Computer Engineering, University of California, San Diego, 9500 Gilman Dr., La Jolla, California 92023, USA}

% To be edited by editor
%\dates{Compiled \today}

% To be edited by editor
% \doi{\url{http://dx.doi.org/10.1364/optica.XX.XXXXXX}}

\begin{abstract}
Exploring and controlling the physical factors that determine the topography of thin liquid dielectric films are of interest in manifold fields of research in physics, applied mathematics, and engineering and have been a key aspect of many technological advancements. Visualization of thin liquid dielectric film topography and local thickness measurements are essential tools for characterizing and interpreting the underlying processes. However, achieving high sensitivity with respect to subnanometric changes in thickness via standard optical methods is challenging. We propose a combined imaging and optical patterning projection platform that is capable of optically inducing dynamical flows in thin liquid dielectric films and plasmonically resolving the resulting changes in topography and thickness. In particular, we employ the thermocapillary effect in fluids as a novel heat-based method to tune plasmonic resonances and visualize dynamical processes in thin liquid dielectric films. The presented results indicate that light-induced thermocapillary flows can form and translate droplets and create indentation patterns on demand in thin liquid dielectric films of subwavelength thickness and that plasmonic microscopy can image these fluid dynamical processes with a subnanometer sensitivity along the vertical direction.
\end{abstract}

\maketitle

\section{Introduction}

Determining the topography of TLD films is of fundamental importance for the basic studies of interfacial science such as the study of wettability and spreading \cite{craster2009dynamics}, 
%\cite{bonn2009wetting,craster2009dynamics}
the study of the response to external stimuli in both physical \cite{oron1997long} and biological systems \cite{grotberg2001respiratory}, and studies related to numerous industrial applications such as coatings, insulating layers and surface modifiers \citep{yerushalmi1994suppression,senaratne2005self}.
While one of the commonly used methods to determine the thickness of TLD films is white-light interferometry, the sensitivity of this method to local small thickness variations is low, especially when the film thickness is less than the optical wavelength due to the low reflection from the film's surface. 
One attractive possibility to enhance this sensitivity is to 
utilize strong coupling between collective oscillations of electrons in the metal (surface plasmons) to 
the resulting radiated electromagnetic field (the polariton), which supports surface plasmon polaritons (SPPs) - electromagnetic waves that propagate along the metal-dielectric interface and strongly decay in the direction perpendicular to it
\cite{raether2013surface}.
% \cite{raether2013surface,maier2007plasmonics}. 
This localization property makes SPPs very sensitive to the dielectric properties of materials in close proximity to the metal surface, which, over the last few decades, has led to a wide spectrum of sensing and imaging techniques, such as surface plasmon spectroscopy and surface-plasmon resonance microscopy (SPRM)  \cite{rothenhausler1988surface}.
% \cite{rothenhausler1988surface,hickel1989surface,flatgen1995two}.
In particular, SPRM is a well-established method for label-free imaging of low contrast features such as the roughness of thin solid dielectric films \cite{sadowski1995characterization},
imaging of small biological objects such as bacteria, viruses, and biofilms, and the detection of analytes as a result of surface reactions. The latter typically results in a few nanometer-thick monolayer array of molecules bound to metal surfaces, which modifies the SPP momentum (e.g., by changing the refractive index), and numerous applications such as the biosensing of surface reaction events 
\cite{lofas1991bioanalysis,homola2006surface}
%\cite{lofas1991bioanalysis,yeatman1996resolution,homola2006surface,campbell2007spr}, 
and cell adhesion sites \cite{watanabe2012scanning}, have been found. 
%cells adhesion and cell based diagnosis \cite{peterson2009surface, watanabe2012scanning,yanase2014surface}. 
Unlike alternative methods suited to 
characterizing nanometric and subnanometric topographies, such as atomic force microscopy (AFM), which are scanning imaging techniques, SPRM is an optical technique that can flexibly measure topographies in both wide-field imaging and scanning confocal modes. Moreover, SPRM is a noncontact method that avoids the sample-specific mechanical interactions involved in AFM imaging. In particular, for fluid topometry (i.e., topography measurement), noncontact optical imaging avoids sensitive perturbations to the TLD dynamics that would otherwise be inevitable with contact mechanical probing.

\sloppy

\begin{figure*}[t]
	\includegraphics[scale=0.21]{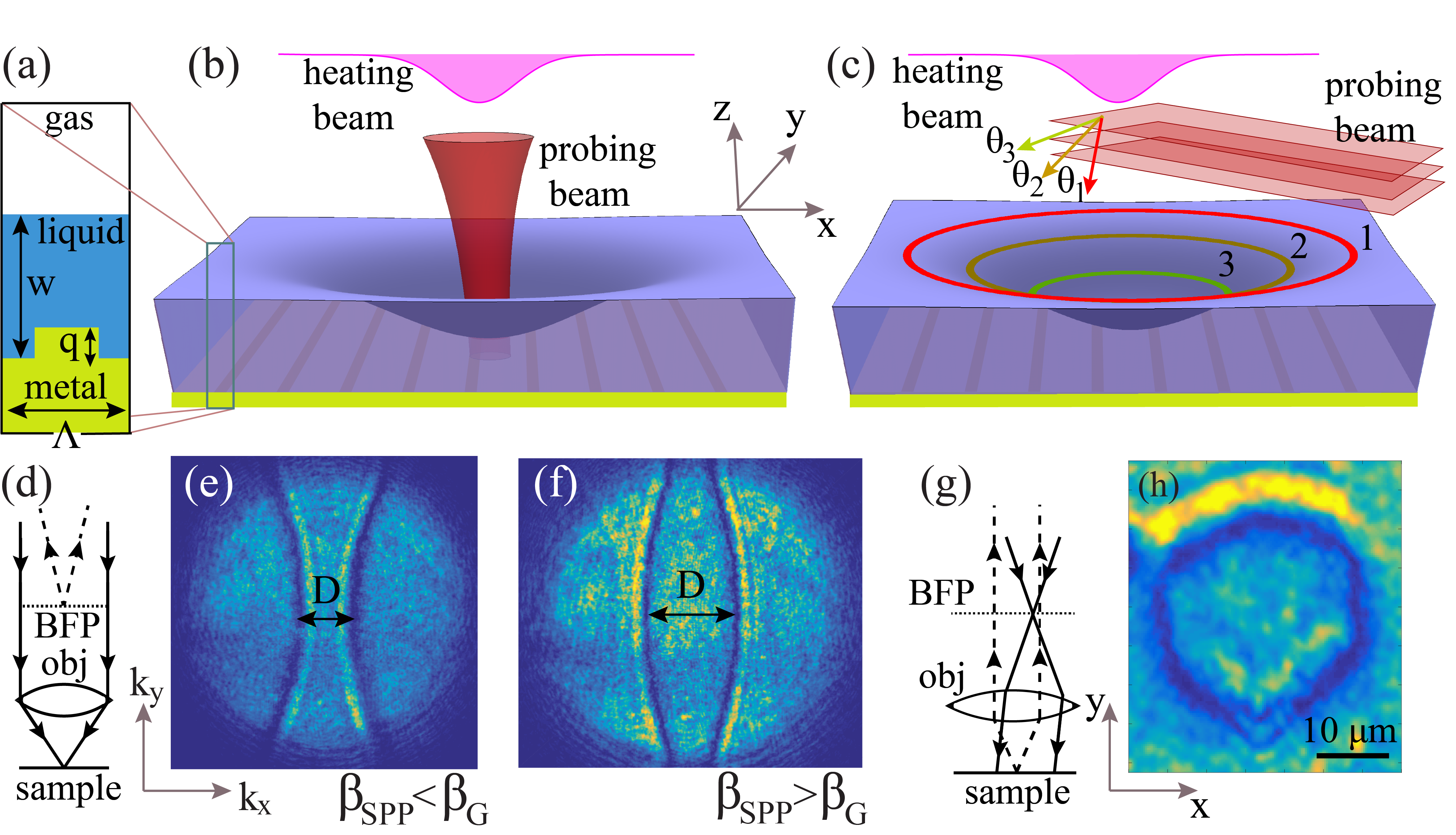}
    \caption{(a) The system under study; a metal grating of period $\Lambda$ and depth $q$ covered with a TLD film of local thickness $w$. (b) $k$-space method: the probing focused Gaussian beam illuminates a small (diffraction limited) spatial region. (c) Real-space method: the probing beam is a plane wave of fixed directionality that covers a wide field of view of several tens of microns. (d,g) The corresponding schemes highlighting the key differences between the topometry methods. (e,f) Experimental representative images of the BFP with dark arcs indicative of the resonant coupling angles to SPP for the regimes $\beta_{SPP}<\beta_{G}$ (e) and $\beta_{SPP}>\beta_{G}$ (f), respectively. (h) Experimental representative image of a resonant contour due to an incident beam at some angle, which corresponds to a unique film thickness. By sweeping the angle of incidence ($\theta_{1,2,3}$ in (c)) of a given light source, the coupling to the SPP takes place at corresponding heights (see contours $1,2,3$ in (c)).}
    \label{Setup}
\end{figure*}

In this work, we demonstrate for the first time a combined imaging and optical patterning projection platform, that is capable of optically inducing dynamical processes in TLD films and then measure subnanometric topographic and thickness changes. Specifically, the optically projected pattern induces thermal gradients which in turn trigger thermocapillary (TC) flows, while SPRM imaging leverages the near-field profile of surface plasmon resonances to resolve and characterize the sub-nanometric topographies of flow-induced interface perturbations. 
Analogous to traditional pump-probe systems, our system uses a heating pump beam to invoke TC flows and is capable of probing the accompanying deformation of the TLD film. 

In our work we introduce local changes in TLD film thickness by taking advantage of the characteristic optical absorption in metals (i.e., unrelated to the SPP-driven heating), which stimulates the TC effect \cite{levich1962physicochemical}. The latter is a special case of the Marangoni effect, 
%\cite{marangoni1871}, 
which is manifested by TC flows and the deformation of a TLD film due to the spatially non-uniform temperature distribution of the liquid-gas interface and subsequent spatial variations of the surface tension. 
Notably, in previous works, heat-induced tuning of the SPP coupling condition was mostly invoked by thermo-optic and phase transition effects (see \cite{vivekchand2012liquid} and references therein); in our work, heat-generated SPP tuning is experimentally triggered for the first time by a fundamentally different mechanism that stems from geometrical changes of the TLD film topography, as was recently theoretically demonstrated \cite{rubin2018nonlocal}. 

Previous studies that applied SPRM imaging to thin dielectric films, such as \cite{huang2007surface}, 
%\cite{nelson1999near,huang2007surface}, 
focused on solid and static films and did not employ the SPRM method to study fluid dynamical processes. 
%In particular, while the TC effect has been extensively studied over the last few decades both experimentally \cite{cazabat1990fingering,singer2017thermocapillary} and theoretically \cite{oron1997long}, not sufficient attention was given to study optically driven TC dynamics on a sub-nanometer scale and dynamical patterning of TLD films.
Furthermore, while TC-induced patterning has received significant attention in the past,
mostly in the context of thin polymer film molding, where TC flows are triggered only in regions where the polymer film is above the glass phase transition temperature \cite{singer2016focused} (and also \cite{singer2017thermocapillary} and references therein), we here leverage the ability to induce optically desired heating patterns to demonstrate the formation of light-induced droplets of different size directly from TLD film and their translation along the substrate
 (see also \cite{yoo2015directed} and \cite{han2017controllable}, which report the formation of droplets from amorphous silicon and gold by employing phase plates). To the best of our knowledge, previous studies in micro- and nanofluidics reported controlled droplet formation in microchannels by employing directional flows of several liquid phases (see \cite{zhu2017passive} for a recent review and references therein) and by utilizing a light-driven TC effect \cite{baroud2007thermocapillary}; however, these studies did not employ optical patterns to take advantage of the TC effect for the formation and translation of microdroplets directly from stationary TLD films without microchannels. 

We experimentally employ traditional SPRM to image the dynamical depth changes and 3D nanoscale topography of the TLD film in a subwavelength thickness regime, which is challenging for purely interferometric methods. A numerical model is constructed that relates TLD thickness to the measured SPP coupling angle, allowing for the experimental determination of the TLD topography and local thickness from a series of SPRM frames. While in our work we experimentally determined the topography of TLD films with a mean thickness comparable to or smaller than the penetration depth of an SPP mode in the direction normal to the metal surface, we theoretically note that in principle, it is possible to achieve similar sensitivity for subnanometer thickness changes of significantly thicker TLD films by leveraging waveguide (WG) modes, which admit the highest optical intensity within the film. 

To illustrate the optical interaction with the TLD film for both manipulation and imaging, Fig. \ref{Setup}(a) presents key elements of our system under study: a metal grating of period $\Lambda$ and wave vector $\beta_{G} \equiv 2 \pi / \Lambda$ and an adjacent layer of the TLD film of thickness $w$, as measured from the bottom part of the grating grooves. Our imaging methods rely on capturing the angular and spatial content of the beam reflected from the sample, referred to below as $k-$space and real-space imaging, which correspond to the monochromatic coupling illumination configurations 
described in Fig. \ref{Setup}(b,d) and Fig. \ref{Setup}(c,g), respectively. $k-$space imaging utilizes a probing focused Gaussian beam, produced by a collimated beam filling the objective back focal plane (BFP), which corresponds to a focused beam on the sample (Fig. \ref{Setup}(b,d)). 
Real-space imaging on the other hand, utilizes the dual configuration with a plane wave made incident at a specified angle on the sample (Fig. \ref{Setup}(c)), which corresponds to a focused beam at the BFP (Fig.\ref{Setup}(g)). Both methods take advantage of the fact that the magnitude of the corresponding SPP momentum, $\beta_{SPP}$, is a strictly increasing function of the TLD film thickness which allows the construction of a one-to-one correspondence between the film thickness and the SPP resonant coupling angle. 

Fig. \ref{Setup}(e,f) presents representative $k-$space imaging results of the BFP with inward and outward facing dark arcs separated by a distance $D$ along the central line, which correspond to different SPP coupling regimes depending on the sign of the difference $\beta_{SPP}-\beta_{G}$.
Importantly, the distance $D$, which in our work can depend on time, together with the inward or outward pointing configuration of the dark arcs, uniquely sets the SPP coupling angle and the corresponding TLD film thickness at the focal point (see \cite{kvasnivcka2014convenient} for BFP imaging due to reflectance from a metal grating covered with a solid dielectric film). The emergent patterns are related to the Kikuchi patterns 
%\cite{kikuchi1928diffraction} 
or Kossel lines
%\cite{kossel1935richtungsverteilung}
 that have been observed in atomic structures through electron microscopy and Brillouin zone spectroscopy of photonic crystals \cite{bartal2005brillouin}.
In the real-space imaging setup (Fig. \ref{Setup}(c,f)), the regions of low reflected light are indicative of spatial regions 
where the resonant coupling condition of the incident plane wave into the SPP mode holds. Fig. \ref{Setup}(h) presents a representative image of a wide field of view of a few tens of microns showing a dark contour indicative of fluid thickness that supports the SPP coupling condition at the illumination angle. 
By sweeping the incidence angle with the same source wavelength, one can then obtain a set of resonant curves that correspond to a set of level thickness contours where the incident plane wave couples to the SPP, which is schematically shown in Fig. \ref{Setup}(c). 

\section{Governing equations}

\subsection*{Thin liquid film deformation due to optically induced thermocapillary effect}

The topography of a TLD film is determined by the internal stress balance between capillary and viscous forces, as well as by external stimuli. Assuming a Newtonian fluid of viscosity $\mu$ (which holds for the silicone oil employed in our work), considering linear dependence of the surface tension, $\sigma$, on the temperature difference $\Delta T$ \cite{levich1962physicochemical} given by
\begin{equation}
	\sigma(T)=\sigma_{0}-\sigma_{T} \Delta T; \quad \Delta T \equiv T-T_{0},
\label{SurfaceTensionGrad}
\end{equation}
and applying the lubrication approximation %\cite{howison2005practical} 
for the corresponding linearized Navier-Stokes equation \cite{oron1997long} yields the following equation for the TLD film deformation $\eta$ under spatiotemporal-dependent optical intensity $I$,
\begin{equation}
	\dfrac{\partial \eta}{\partial t} + \nabla^{4}_{\parallel}  \eta =  - \text{Ma} \cdot  \chi \cdot \frac{\tau_{l}}{\tau_{th}}  I/2
\label{ThinFilmQuasiStatNonDim}    
\end{equation}
(see \cite{rubin2018nonlocal} and Supplementary Material).
Here, $\sigma_{T}$ is the so-called Marangoni constant; $\text{Ma} \equiv \sigma_{T} \Delta T w_{0}/(\mu D_{th}^{m})$ is the dimensionless Marangoni number, which represents the ratio between the surface tension stresses due to the TC effect and dissipative forces due to fluid viscosity and thermal diffusivity in a film of height $w_{0}$; $\chi \equiv (\alpha_{th}^{m}  d^{2} I_{0})/(k_{th}^{m} \Delta T)$ is the dimensionless intensity of the heat source; $d$ is the typical length scale along the in-plane direction; $I_{0}$ is the characteristic optical power scale; $D_{th}^{m} \equiv k_{th}^{m}/(\rho^{m} c_{p}^{m})$ is the heat diffusion coefficient;  $\rho^{m}$, $c_{p}^{m}$, $k_{th}^{m}$, and $\alpha_{th}^{m}$ are the mass density, specific heat,
heat conductance, and optical absorption coefficient of the metal substrate, respectively; 
$\tau_{th} \equiv d^{2}/D^{m}_{th}$ is the characteristic heat diffusion time scale in the metal; $\tau_{l} \equiv d^{4}/D_{\sigma}$ is the typical thin film deformation time scale; and $D_{\sigma} \equiv \sigma_{0} h_{0}^{3}/(3 \mu)$.
Importantly, the negative (positive) sign of the source term in Eq.(\ref{ThinFilmQuasiStatNonDim}) indicates that local regions of higher (lower) temperature lead to a local decrease (increase) in TLD film thickness. 

\subsection*{Coupling of SPP into grating covered with dielectric}

Grating coupling is a common method to evanescently excite SPPs from free-space modes (see \cite{homola2006surface} and references therein). Assuming that a plane wave with a wave vector component parallel to the grating, $\vec{\beta}_{I}$, is made incident on a metal surface with a grating of period $\Lambda$ gives rise to a series of diffracted waves with wave vectors given by $\vec{\beta}_{I}+N\vec{\beta}_{G}$, where $\vec{\beta}_{G}=\left( 2 \pi/\Lambda \right) \hat{k}_{x}$ is the grating vector that is perpendicular to the grating grooves (aligned along the $y$ direction as described in Fig. \ref{Setup}) and $N$ is the diffraction order. The diffracted waves along the interface can couple to an SPP mode provided the following momentum balance equation holds \cite{raether2013surface}:
\begin{equation}
	\beta_{I} + N\beta_{G} =  \text{Re}\left( \beta_{SPP} \right),
\label{MomentumConserv}    
\end{equation}
where $\beta_{I}=k_{0}\sin(\theta)$, and $\beta_{SPP} = k_{0} \sqrt{\epsilon_{m}\epsilon_{d}/(\epsilon_{m}+\epsilon_{d})}$ is the SPP momentum \cite{raether2013surface}. Here, $\theta$ is the resonance coupling angle of incoming light, $k_{0}=2 \pi / \lambda$ is the magnitude of a free-space wavenumber vector of wavelength $\lambda$, and $\epsilon_{m}$ and $\epsilon_{d}$ denote the dielectric constants of the metal and dielectric, respectively. 
Inserting the definitions of $\beta_{I}$, $\beta_{G}$ and $\beta_{SPP}$ into Eq.(\ref{MomentumConserv}) yields the following expression for $\sin(\theta)$:
\begin{equation}
	\sin(\theta) = \sqrt{\dfrac{\epsilon_{m}}{\epsilon_{m}+n_{d}^{2}}} - \dfrac{\lambda N}{\Lambda n_{d}},
\label{sintheta}	
\end{equation}
where $n_{d}=\sqrt{\epsilon_{d}}$.
Treating the TLD film and air bilayer as a single medium with a depth-averaged index, obtained by integrating the index distribution along the direction normal to the metal surface \cite{campbell1998quantitative}, we learn that the resonant coupling angle is a strictly decreasing function of the TLD film thickness. In particular, as the TLD film thickness increases in the range $0 \leq w < w_{c}$, the resonant coupling angle tends to a normal incidence condition at critical thickness $w_{c}$. For higher thickness values, i.e., for $w>w_{c}$, the coupling angle becomes negative, which corresponds to a transition from the counter-propagating to the co-propagating regime of the coupled SPP mode (see Fig.S1 in Supplementary Material). 
To obtain a quantitative model that predicts the TLD film thickness (without approximations such as the depth averaging mentioned above) based on an experimentally measured resonant coupling angle, we turn to a numerical simulation (using Lumerical FDTD). While in real-space imaging the resonant coupling angle coincides with the predetermined illumination angle, in $k$-space imaging, the resonant coupling angle can be inferred from the distance between the dark arcs, $D$, shown in the representative experimental images of Fig. \ref{Setup}(e,f), via
\begin{equation}
	\theta=\sin^{-1}(D/2k_{0}),
\end{equation}
where $D$ and $\theta$ can be functions of time due to thickness changes of the TLD film.

\subsection*{Experimental setup}

We employ two different laser sources, the so-called heating- and probing beams described in Fig. \ref{Setup}, which are coupled to the same optical path, and deliver them to the sample. An argon laser ($\lambda =488$, $514$ nm) is used as the heating beam, which heats the metal substrate by optical absorption of the projected pattern and invokes TC flows accompanied by deformation of the TLD film; an NIR laser diode ($\lambda=785$ nm) is used as the probing beam, which, upon reflection provides an angular and spatial map for resonant SPP coupling.

For a case of large deformations of the TLD film, which are within the reach of standard white-light microscopy, we employed a similar experimental setup, with a $\lambda=532$ nm heating beam. An epi collection microscope was used to deliver both beams and image the fluid response (Olympus). The heating beam was collimated, optionally passed through a beam-shaping or transparency element, and then passed through a $50$X
microscope objective (Mitutoyo) and focused to a point with a $1.5$ $\mu$m width (FWHM). 
The probing beam was delivered to the microscope objective in one of two configurations: collinear with the pump beam as a plane wave and focused on the sample (real-space imaging) and focused on the objective BFP and delivered to the sample as a plane wave (k-space imaging), both described in 
Fig. \ref{Setup}(d) and Fig. \ref{Setup}(g), respectively. 
In the first configuration, the probing-beam was then epi collected, and the BFP was imaged to measure the angular spectrum up to the objective numerical aperture, where the input angles satisfying the SPP momentum conditions appear as absorption lines in the angular spectrum. 
Changes in the momentum condition owing to the thickness change due to TC and healing flows were visualized as shifts in the resonance angle.  
In the second configuration, the collimated probing beam was passed through a lens (f = $40$ cm) on a lateral translation stage and focused onto the BFP for K\"{o}hler illumination. Lateral displacements of the lens shifted the focus across the BFP, thus shifting the plane wave angle of incidence at the grating. The collected light was imaged by the microscope tube lens for direct observation of the grating surface, and regions with TLD film thickness matching the SPP condition at a given angle of incidence could be observed to darken according to resonant absorption. Because of the large thickness range in the fluid layer after the TC flow, sweeping of the illumination angles allowed visualization of the resonant contours of the corresponding level regions with thicknesses supporting the SPR resonance at the given angle. To form the TLD film, silicone oil of refractive index $1.39$ was spun onto the fabricated gold grating (see Supplementary Material for more details). 

\section{Model for TLD film thickness determination}

\begin{figure*}[ht]
	\includegraphics[scale=0.35]{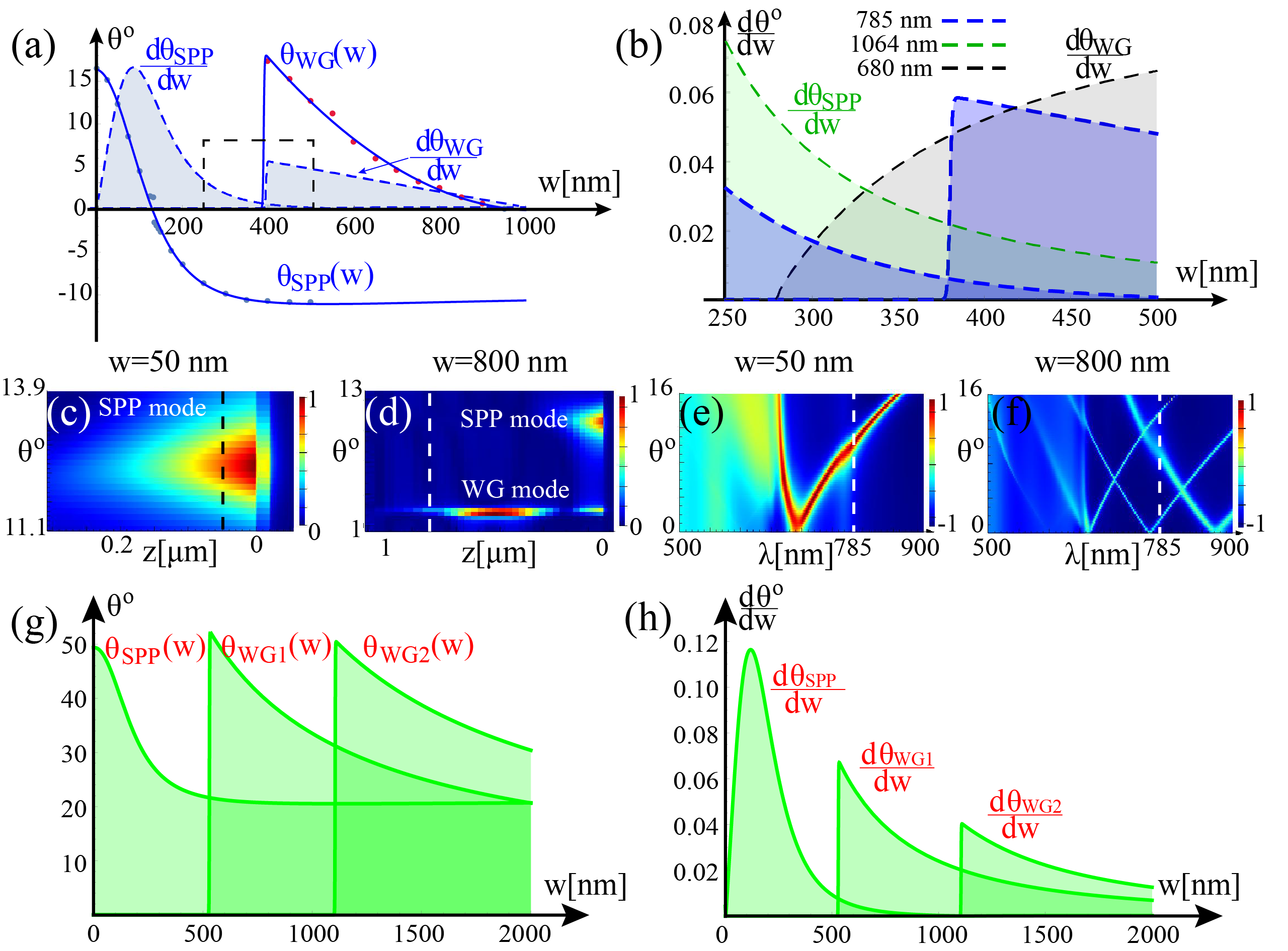}
    \caption{Numerical simulation results that relate the resonant coupling angle $\theta$ to the TLD mean film thickness $w$. (a) Discrete values and smooth interpolating curves, $\theta_{SPP}(w)$ and $\theta_{WG}(w)$, for $\lambda=785$ nm, accompanied by the corresponding normalized sensitivity curves (filled). (b) The sensitivity curves of SPP and WG modes for $\lambda=1064$ nm and $\lambda=680$ nm, respectively, around the region of lower sensitivity for $\lambda = 785$ nm (dashed black box in (a)). (c,d) Electrical field intensity colormap in a $\theta-z$ plane, with dashed lines indicating the corresponding thickness values of $w=50$ nm and thicker $w=800$ nm films. (e,f) The corresponding dispersion relation. (g) The $\theta_{SPP}(w)$, $\theta_{WG1}(w)$, and $\theta_{WG2}(w)$ curves for a $\lambda=1064$ nm probing beam and (h) the corresponding sensitivity curves.}
    \label{SPPandHybridCurves}
\end{figure*}

Extracting the TLD film thickness from experimental measurements requires a theoretical model that relates the thickness of the TLD film to the resonant coupling angle of the substrate guided modes. Fig. \ref{SPPandHybridCurves} presents numerical simulation results of an incident plane wave coupling into propagating SPP and WG modes on a metal grating covered with a TLD film by using Lumerical FDTD.
The key components of the simulation domain, presented schematically in Fig.\ref{Setup}(a), consist of an incoming plane wave of wavelength $\lambda = 785$ nm, grating period $\Lambda = 600$ nm, and grating depth $q=30$ nm. The dielectric constant of gold at $785$ nm and the refractive index of silicone are taken as $\epsilon_{m} = -22.85+1.4245 i$ \cite{johnson1972optical} and $n_{d} = 1.39$, respectively 
(see Supplementary Material for additional details).
Fig. \ref{SPPandHybridCurves}(a) presents the resonant coupling angle curves $\theta_{SPP}(w)$ and $\theta_{WG}(w)$ into an SPP mode and into a 
higher-order WG mode (which emerges at higher thickness values \cite{aust1994surface}), respectively, as a function of the TLD film thickness $w$. 

Importantly, both are strictly decreasing functions of $w$ and therefore can be used to construct an inverse function where $w$ can be deduced based on the values of $\theta$. The regions of positive and negative values of the function $\theta_{SPP}(w)$ correspond, respectively, to SPP modes that propagate in a counter- and co-propagating direction relative to the in-plane component of the incident plane wave.  These two different regimes are separated by an angular band gap centered at the normal incidence angle (see Supplementary Material section), which is reminiscent of the band gap observed near normal incidence conditions on metal gratings without a dielectric layer \cite{pang2007observation} due to destructive interference of the left- and right-propagating SPP modes. The filled curves in Fig. \ref{SPPandHybridCurves}(a) present the corresponding sensitivity for the SPP and the higher-order WG mode, defined as $d \theta_{SPP}(w)/dw$ and $d \theta_{WG}(w)/dw$, respectively. While the SPP mode admits the highest sensitivity value for a film thickness of approximately $100$ nm and practically vanishes for thickness values above $600$ nm, the first WG mode (which does not emerge at low thicknesses) admits the highest sensitivity at approximately $400$ nm. These differences in regions of maximal sensitivity stem from the fact that, while for sufficiently thin films, the incoming plane wave couples to an SPP mode that admits the highest optical intensity on the metal surface (see Fig. \ref{SPPandHybridCurves}(c)), for thicker films the incident plane wave couples into a higher-order WG mode with the highest optical intensity at some finite distance from the metal surface within the TLD film, as presented in Fig. \ref{SPPandHybridCurves}(d) (see also \cite{tobias2007theoretical} for optical intensity distribution along the vertical direction in dielectric-loaded SPP waveguides). Interestingly, the sensitivity of the WG modes drops in close proximity to the cutoff thickness values (i.e., film thickness that satisfies the WG cutoff condition \cite{kogelnik1974scaling}), seen around $w=375$ nm in the center of the dashed region in Fig. \ref{SPPandHybridCurves}(a). 
As presented in Fig.\ref{SPPandHybridCurves}(b) one could potentially mitigate these lower-sensitivity regions by employing either a longer wavelength probing beam which can couple to an SPP mode, or alternatively a shorter wavelength probing beam which can couple to a WG mode, without modifying the grating properties.

%\begin{figure}[b!]

Fig. \ref{SPPandHybridCurves}(e) presents the dispersion relation of the basic SPP mode, which is coupled to the metal grating that hosts a $w=50$ nm dielectric. In particular, it shows a single curve, where the right and left branches correspond, respectively, to counter- and co-propagating modes relative to the in-plane component of the incident plane wave. The dashed vertical line at $\lambda=785$ nm, which corresponds to the wavelength of the probing beam, intersects with the right branch, indicating that the given conditions facilitate coupling only to the counter-propagating SPP mode at a resonant coupling angle of $12.28^{o}$. Since the dispersion relation of WG modes propagating in a slab WG (formed, in our case by the TLD film) is set by the constructive interference condition \cite{kogelnik1974scaling}, it follows that additional higher-order WG modes must emerge at even higher thickness values of the TLD film. Indeed, Fig. \ref{SPPandHybridCurves}(f) presents an additional higher mode that emerges in thicker films of thickness $800$ nm. The dashed vertical line of the probing beam at $\lambda=785$ nm intersects with a left (co-propagating) branch of the SPP curve at $11.12^{o}$ and with the right branch of the first WG mode at $2.51^{o}$. 

Fig. \ref{SPPandHybridCurves}(g) presents resonant coupling angle as a function of film thickness for the SPP mode as well as for the first and the second WG modes, for the case of a longer wavelength ($\lambda=1064$ nm) probing-beam, whereas Fig.\ref{SPPandHybridCurves}(h) presents the corresponding sensitivity curves. Similarly to the case of the $\lambda=785$ nm probing-beam described above, thicknesses around the cutoff condition of each one of the WG modes are accompanied by lower sensitivity regions. As expected, employing higher wavelength allows to extend the sensitivity range for higher film thicknesses due to higher penetration depth of the corresponding SPP mode into the dielectric region; for instance a sensitivity of $0.03^{o}$/nm can be achieved by employing $785$ nm wavelength at thickness $713$ nm or by employing $1064$ nm wavelength at thicknesses $872$ nm or $1291$ nm. It is instructive to mention that in our work SPP and WG modes admits more than an order of magnitude higher sensitivity than the FP (Fabry-Perot) reflectance commonly used in white light microscopy (see Supplementary Material).

\section{Experimental results}

\subsection{Dynamical optical manipulation of TLD films}

\begin{figure}[th]
	\includegraphics[scale=0.45]{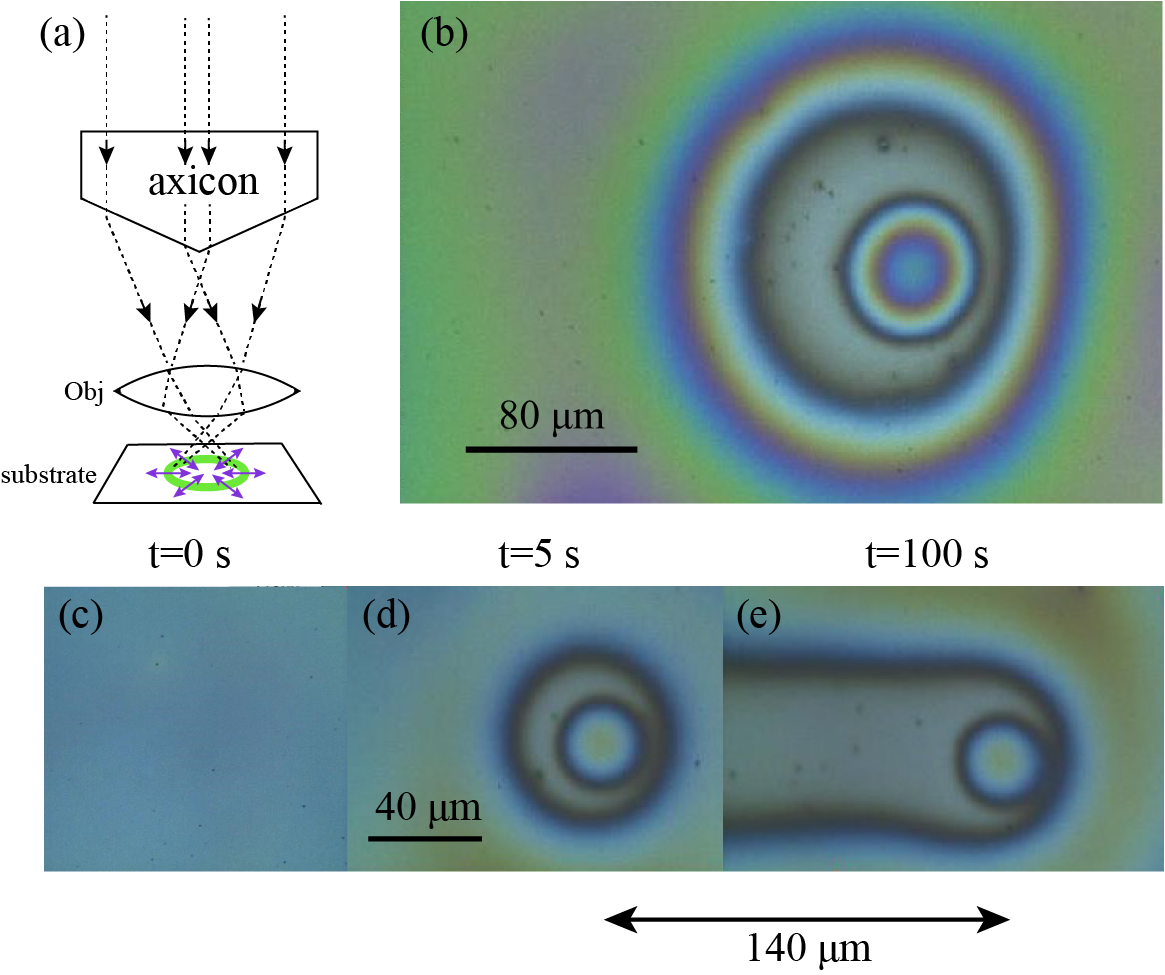}
    \caption{White-light microscopy experimental images of optically driven droplet formation and translation along the substrate. (a) Ray-path scheme presenting refraction from the axicon's conical surface, the ring-shaped illumination pattern on the substrate and the direction of mean TC flow (blue arrows). (b) Optically induced droplet of $60$ $\mu$m in diameter from a TLD film of thickness $400$ nm due to a $400$ mW laser power source of wavelength $532$ nm operating for $30$ s ($20$X objective). The asymmetry of the annular-shaped deweted region stems from slight asymmetry in the illumination pattern. (c,d,e) Time sequence of the formation of a droplet with a diameter of approximately $30$ $\mu$m by applying a similar laser source to a $200$ nm thick TLD film and a $50$X objective; (c) prior to illumination $t=0$ s, (d) $t=5$ s, and (e) $t=100$ s and a $140$ $\mu$m horizontal substrate translation.}
    \label{AxiconDroplet}
\end{figure}

To probe optical control over the shape of the liquid-gas interface, we employed a novel optical patterning system to deliver light illumination through a beam-shaping element. The delivered light pattern induces a thermal gradient on the liquid-gas interface and consequently triggers TC flows, which allow manipulation of its shape. In particular, fixing an axicon beam-shaping element to the Fourier plane of the objective lens transforms the substrate plane focal point into a focused ring-shaped illumination pattern on the substrate (Fig. \ref{AxiconDroplet}(a)) and leads to TC flows that invoke film thinning in the regions of highest temperature. Fig. \ref{AxiconDroplet}(b) presents a white-light microscopy image of the generated droplet of approximately $60$ $\mu$m in diameter from a $400$ nm thick silicone oil film, spun on a NiCr substrate, due to a $400$ mW laser source of wavelength $532$ nm and an illumination time of $30$ s. Film deformations were measured using a white-light reflection microscope (Leica, 50X) and a $20$X objective.
Fig. \ref{AxiconDroplet}(c-e) present droplet formation from a thinner $200$ nm thick silicone oil film and droplet translation along the metal substrate, achieved by translating the microscope stage and keeping all other components of the optical setup stationary. In this case, we utilized a higher magnification objective, $50$X, which resulted in a higher optical power and consequently shorter illumination time of $5$ s. The speed of stage translation in our experiments was approximately $1.4$ $\mu$m/s, which was sufficiently low to ensure that the invoked TC flows in the silicone oil film around the droplet led to a ring-shaped rupture as the optical pattern was progressively translated along the substrate. For a fixed optical power and faster substrate translation rates, the droplet size was progressively reduced. Interestingly, since the healing time of the silicone oil film due to the formed rupture was longer than the translation time of the substrate, a wake behind the droplet was formed. 

While for the analysis of optically thick TLD deformations white-light contrast is sufficient, high-sensitivity characterization is much more challenging for processes where the resultant TLD film thickness variations are small. Below, we apply the SPRM method coupled to our optical maniupation system to demonstrate much higher sensitivity.

\subsection{$k$-space imaging}

\begin{figure*}[th]
	\includegraphics[scale=0.25]{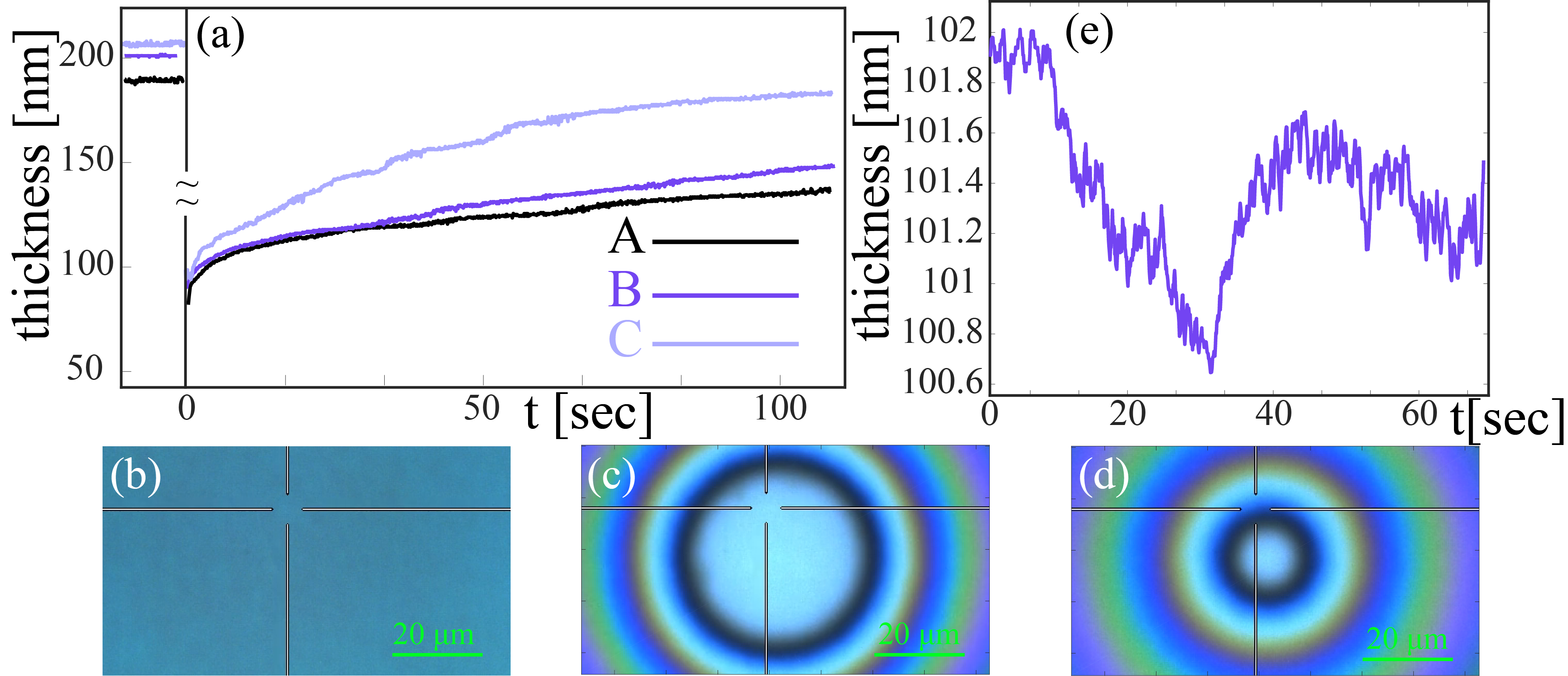}
    \caption{(a) $k$-space imaging: TLD film thickness as a function of time at the center of a focused heating beam (beam waist of diameter $0.62$ $\mu$ m) for three different optical powers: (A) $34.57$ mW, (B) $17.96$ mW and (C) $1.79$ mW; the corresponding intensities on the sample are given by (A) $1.14 \cdot 10^{11}$ W/m$^{2}$, (B) $5.94 \cdot 10^{10}$ W/m$^{2}$ and (C) $5.92 \cdot 10^{9}$ W/m$^{2}$. We attribute the difference in film thickness between different experiments at times $t<0$ to spontaneously formed nonuniformity of the free interface. (b,c,d) Top view of white-light microscopy images of TLD film rupture under optical power employed in case A, captured at three moments in time: (b) before optical heating, (c) $10$ s after the heating is switched off and (d) $30$ s after the heating is switched off. (e) Film thickness as a function of time, on a subnanoscale, due to $20$ s of illumination of a much less intense beam of intensity $1.98 \cdot 10^{7}$ W/m$^{2}$ (power $6$ $\mu$W) and TLD film healing after the beam is switched off at $t=30$ s. 
In particular, the silicone oil thickness change during the period from $30$ s to $35$ s, is from $\sim 100.6$ nm to $\sim 101.2$ nm, which
is approximately $0.6$ nm.    
%    (A) 1.60 37.6 (B) 1.60 27.5 (C) 1.60 17.5 % Spot area 0.62 micron^2
    }
    \label{ZeroAngle}
\end{figure*}

Fig. \ref{ZeroAngle}(a) presents the experimental results of silicone oil film thickness as a function of time at the illumination spot due to three different optical heating powers. Prior to applying the laser heating beam ($t<0$), all three experiments show nearly constant thickness values, whereas at later times ($0 - 20$ s), the fluid thickness is reduced due to the TC flows, which carry the fluid towards regions of lower temperature away from the focused illumination spot. 
As expected, higher optical powers lead to thinner films; however, in cases $A$ and $B$ of highest optical intensity, the corresponding thickness as a function of time shows a very similar behavior. This similarity is indicative of a rupture process in the silicone oil film that exposes disk-shaped regions on the substrate, as shown in the white-light microscopy images in Fig. \ref{ZeroAngle}(b-d), for a similar optical power and thicker ($0.5$ $\mu$m) silicone oil film. 
Specifically, since the rupture expansion rate due to a heating source of fixed intensity progressively decreases as the rupture grows, similar to the thinning rate of optically stimulated, significantly thicker fluid films \cite{wedershoven2014infrared}, we expect to obtain similar ruptures in cases in which the total illumination time is significantly larger than the rupture formation time. After the heating-beam is switched off, all curves $A-C$ show an increase in film thickness, which is indicative of the healing process due to the capillary forces that eventually lead to a nearly flat film.

Fig. \ref{ZeroAngle}(e) presents the thickness measurement of silicone oil at much lower optical power ($6$ $\mu$W), which allows the capture of its dynamics during the operation of the heating beam. Specifically, during the operation of the heating beam from $10-30$ s, the silicone oil film shows a decrease in its thickness, whereas at later times after the heating beam is switched off ($t>30$ s), it shows a healing process characterized by an increase in its thickness, qualitatively similar to the healing process occurring at higher powers described in Fig. \ref{ZeroAngle}(a). 
Notably, the thickness change of the silicone oil film occurs at a subnanometer scale but still over a long period of time, which is on the order of seconds. 
This large characteristic healing time scale allows ruling out other thermally driven effects, such as the thermo-optical effect, which is governed by heat diffusion processes and therefore operates on a much shorter time scale on the order of magnitude of several microseconds (see Supplementary Material). 
Moreover, even for the case where the thermo-optical effect has a nonnegligible effect, it would still be much smaller than the corresponding contribution due to TLD film deformation driven by the TC effect under identical optical power \cite{rubin2018nonlocal}.

\subsection{Real-space imaging}

\begin{figure}[th] % width=\linewidth
	\includegraphics[scale=0.065]{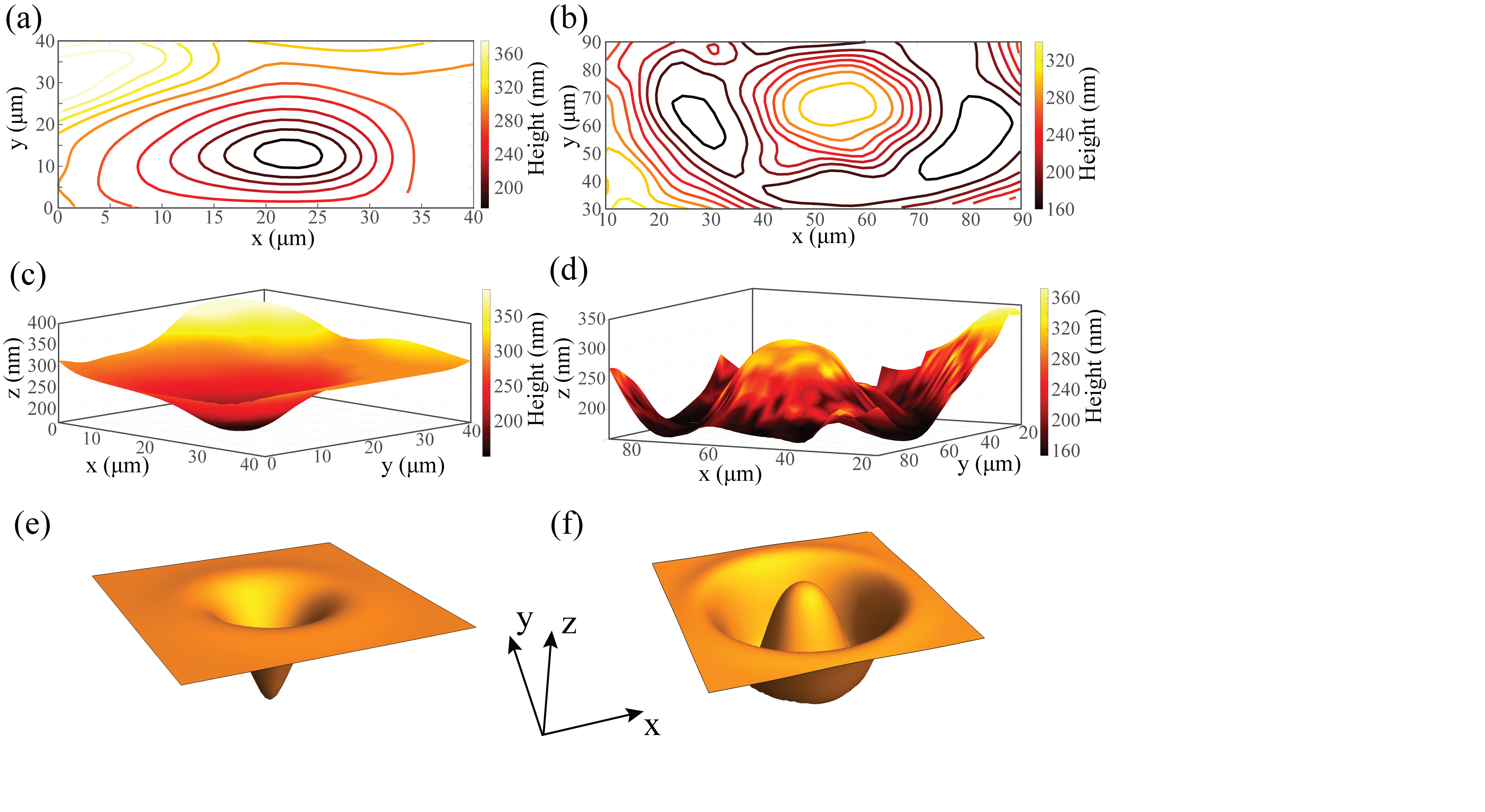}
	\centering
    \caption{Real-space imaging experimental results presenting contour maps (a,b) and interpolated 3D topography maps (c,d) of silicone oil film due to focused (a,c) and ring-shaped (b,d) illumination patterns. (e,f) Numerical simulation results of TLD film deformation Eq.(\ref{ThinFilmQuasiStatNonDim}) for focused (e) and ring-shaped (f) illumination patterns. The size of the domain along the in-plane $y-z$ directions is $20 \times 20$ and the maximal value of the deformation is $0.06$ in dimensionless units (see more details in Supplementary Material).}
    \label{Contour}
\end{figure}

Fig.\ref{Contour} presents the experimental results of TLD film equal-height contour lines as well as the corresponding interpolated topography maps. In particular, it shows an indentation formed under focused illumination in a small region with a diameter of approximately $0.65$ $\mu$m (Fig. \ref{Contour}(a,c)) and a droplet formed under a ring-shaped region with an inner diameter of approximately $40$ $\mu$m (Fig. \ref{Contour}(b,d)). These optically induced heat patterns, which lead to different fluidic patterns, are qualitatively captured by numerical simulation of the corresponding cases presented in 
Fig. \ref{Contour}(e,f).
The dark resonant contours presented in Fig. \ref{Contour}(a,b) correspond to different equal-height contours of the TLD film free surface, obtained by sweeping the corresponding incident angles and capturing the reflected light from the sample. The step in the illumination angle sweep is $0.77^{o}$ per step, and then for each value of the resonant coupling angle we used the theoretical curve to determine the thickness of the silicone oil film. 
%We applied Gaussian smoothing filter of $1.5$ pixels ($8.6$ \textmu m) to all images to remove the noisy features in the contours. 
 
After the heating beam is switched off, the thin film healing rate is relatively fast, as could also be
observed via $k$-space imaging in Fig. \ref{ZeroAngle}(a) just after $t=0$ s. The latter implies that real-space imaging of dynamical processes is applicable to processes on time scales smaller than the time scale required to scan the relevant angles. In the present setup, the scanning time was $10$ s; this consideration dictated performing the angle scan at later times, characterized by a slower healing process, where the droplet already coalesced with the surrounding thin film. 

It is instructive to mention that the curvature of the TLD film in our experiments was sufficiently low, and therefore, its effect on the coupling angle is negligible. Indeed, as seen from Fig. \ref{Contour}(c), the mean slope is approximately $\Delta z/\Delta x = 100 \text{nm}/ 20$ $\mu$m $= 5 \cdot 10^{-3}$, which is smaller than the critical value $8 \cdot 10^{-2}$ predicted by the analysis from Fig. S2(c,d) (see Supplemental Material for details). 

\subsection{Effect of thin film periodic undulation on the coupling angle}

\begin{figure}[th] % width=\linewidth
	\includegraphics[scale=0.25]{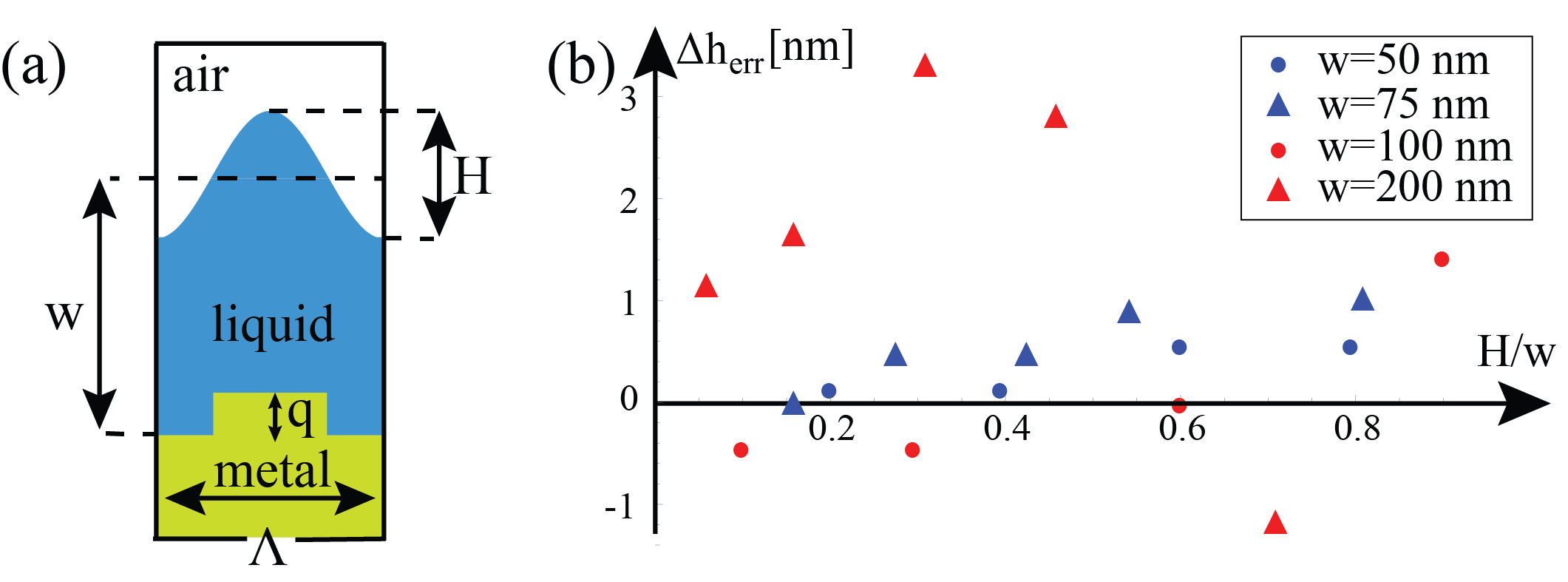}
    \caption{(a) Numerical simulation domain presenting a basic unit cell of period $\Lambda = 600$ nm, and grating depth $q=30$ nm that hosts a symmetrically undulated TLD film of the same period, with peak-to-peak amplitude $H$ and mean thickness $w$. (b) Simulation results of the predicted error of the film thickness, $\Delta h_{\text{err}}$, relative to a flat TLD film with the same mean thickness and $H=0$. In particular (b) presents $\Delta h_{\text{err}}$ as a function of the normalized deformation $H/w$ for mean thicknesses $w=50, 75, 100, 200$ nm.  
%    with the corresponding zones of reduced sensitivity just below the waveguide cutoff conditions around $520$ nm and $1100$ nm}
}
    \label{Undulation}
\end{figure}

We now analyze the effect of possible periodic undulation of the TLD film surface due to an uneven subwavelength coupling grating (see Fig. \ref{Undulation}(a)) on the resonant coupling angle. The governing equation for the thickness of a thin Newtonian liquid film (note that the silicone oil employed in our experiments is a Newtonian fluid) under the effects of surface tension and centrifugal forces due to rotation with angular velocity $\omega$ is given by 
\cite{stillwagon1990leveling}
\begin{equation}
	\dfrac{\partial^{3} \eta}{\partial x^{3}} + \Omega^{2} \Big[ 1 - \left( \frac{w}{w-q} \right)^{-3} + 3 \left( \frac{w}{w-q} \right)^{-4} \eta  \Big] = 0,
\label{NoddomOmega}
\end{equation}
where $\Omega^{2} \equiv  \rho \omega^{2} (\Lambda/2)^{3}  r_{0}/(\sigma (w-q))$, and a similar equation with $q=0$ holds over grooves.
Here, $w-q$ represents film thicknesses above a grating of depth $q$, $r_{0}$ is the distance from the rotation axis, and $\Omega^{2}$ represents a dimensionless ratio of centrifugal to capillary forces. Inserting the following parameters relevant to our system: $\rho=970$ kg m$^{-3}$, $\omega=100$ s$^{-1}$, $\Lambda=6 \cdot 10^{-7}$ m, $r_{0}=10^{-3}$ m, $\sigma=10^{-3}$ N m$^{-1}$, and $w-q=5 \cdot 10^{-8}$ m, into the definition of $\Omega^{2}$ given by Eq.(\ref{NoddomOmega}), we learn that $\Omega^{2} = 2.06 \cdot 10^{-4} \ll 1$. The latter implies that the conditions in our system correspond to a surface-tension-dominated regime, which leads to a practically planar fluid-gas interface (see \cite{wu1999complete} for a case where $\Omega^{2} \approx 10^{-2}$).  Indeed, we find that TLD films of thickness values in the range between $w=50$ nm and $270$ nm acquire a peak-to-peak undulation depth $H$, on the order of $10^{-5}-10^{-7}$ in units of $w-q$, i.e., much smaller than $1$ nm (see Supplementary Material for more details).
%i.e. of the order $10^{-12}-10^{-14}$ m. 
Nevertheless, we note that higher values of $H$ can, in principle, emerge under different experimental conditions that are not met in our work, such as higher values of $\Omega^{2}$ or cases where non-Newtonian fluids or complex fluids with volatile components (e.g. polymers or photoresists) are employed \cite{rawlings2015accurate}. 
%In particular, evaporation of the volatile component leads to increased viscosity and lower surface tension, and can be employed to control roughness of spin cast dielectric films \cite{rawlings2015accurate}, but is not expected to be relevant in the non-volatile silicone oil employed in our work.
%\cite{denis2002fabrication,rawlings2015accurate}. 
Fig. \ref{Undulation}(b) presents numerical simulation results of the error in the TLD film thicknesses $w=50, 75, 100, 200$ nm due to the presented values of the undulation peak-to-peak amplitude $H$ (in units of $w$). Interestingly, even large values of $H/w$ lead to a very small shift in the resonant coupling angle. 
For instance, for the case where $w=50$ nm and $H=30$ nm (which is $60 \%$ of the mean thickness), an error is introduced around $\Delta h_{\text{err}} = 0.5$ nm, which corresponds to only $1.5 \%$ of the mean thickness $w$ (see Supplemental Material for more details). Importantly, combining the small error introduced by the symmetric undulation amplitude $H$ with the analysis above indicating that $H$ is expected to be smaller than one nm, we learn that the corresponding value of $\Delta h_{\text{err}}$ is expected to be much smaller than one nm even in the presence of a $30$ nm deep grating.

\section{Discussion}

In this work we presented dynamical optical patterning of a TLD film and two novel plasmonic methods for the measurement of its thickness and topography, 
which were applied to determine the film dynamics after being driven out of equilibrium by an optically induced  TC effect. 
In particular, we leveraged the TC effect as a novel heat-based mechanism to induce tuning of the plasmonic resonances.
The imaging methods presented in our work are complementary in the sense that while $k$-space imaging provides thickness values as a function of time in a small region, real-space imaging is applicable to a larger area of a few hundreds of microns. Both methods admit immediate generalizations that can be utilized to probe specific fluidic processes by introducing higher-speed sweeping mechanisms (e.g., translation stages) and higher frame-rate cameras; $k$-space imaging can be extended to provide thickness information along spatial segments that can cover 2D regions, whereas real-space imaging can, in principle, capture  faster processes that evolve over smaller characteristic times. Our methods open a door for future studies where thickness changes of the TLD film monitored at a subnanometer scale would be indicative of various physical processes, such as thermocapillary instabilities in polymer molten films \cite {mcleod2011experimental} and instabilities due to electrohydrodynamic \cite{schaffer2001electrohydrodynamic}
%\cite{herminghaus1999dynamical,schaffer2001electrohydrodynamic} 
and electrostatic \cite{morariu2003hierarchial} effects. %dispersion forces \cite{reiter1999thin}, and chemical changes such as light-driven photoisomerization from trans- to cis-isomer states \cite{rochon1995optically}.
%%and heat-transfer during evaporation from non-uniform films \cite{liu1994image}.
In addition, the methods are applicable to studying pore nucleation \cite{ilton2016direct}, phase transitions in thin polymer films \cite{wallace1995influence} and light-induced reversible thickness changes of photosensitive materials \cite{yager2006photomechanical} without the need to employ, respectively, X-ray reflectivity and neutron reflectometry methods. Importantly, our methods are not compromised if an additional thin layer of solid dielectric is deposited on top of the metal grating to achieve desired surface properties, which, in principle, allows studying basic problems such as the wetting and spreading of thin films on different substrates. 
Our analysis of the effect of fluid-gas interface periodic undulation on the resonant coupling angle, due to possible interaction with a nonuniform topography (metal grating), showed that it introduces a negligible error in the predicted thickness values. Nevertheless, periodic undulation could be of importance in future studies involving non-Newtonian fluids with volatile components (e.g., polymers) where the peak-to-peak undulation depth may be larger and asymmetrical. Furthermore, we demonstrated a novel optical method/process to create and translate droplets directly from a TLD film, which merits further investigation from a fluid dynamical perspective and may also be useful for future applications in microscopy/nanoscopy 
\cite{lee2009near}, where each droplet can serve as a smooth, tunable and mobile micro/nanolens, and in bio-oriented applications in microfluidics.
%\cite{lee2009near,wang2011optical,krivitsky2013locomotion}
In fact, since the temperature increase required to form the droplets from the liquid is expected to be only a few degrees or less (see Supplemental Material), we expect that our method may be suitable for various applications in the biosciences, such as single virus communication and PCR (polymerase chain reaction) assays, where compartmentalization of the TLD film into an array of droplets may be useful to invoke independent biomolecular reactions \cite{kelley2014advancing}.
We hope that our work, which brings together plasmonic effects, SPRM and fluid dynamics, will stimulate future development of specific optical systems aimed at studying TLD film processes and light-fluid interactions. In particular, other prominent SPRM variants that can be employed include the classical Kretschmann configuration to study the dynamics of TLD films of thicknesses below the grating depth, the holographic-based method \cite{mandracchia2015surface}, which offers the possibility of simultaneous detection of reflectivity and phase changes in SPR images, and SPP confocal interferometry \cite{pechprasarn2014ultrastable}, which offers superior noise performance and can improve the lateral resolution.

\section*{Materials and Methods}

For the TLD film substrate which supports SPP excitations we emloyed gold gratings with a period of $600$ nm, $50$\% duty cycle, and with $30$ nm depth grooves which were fabricated by nanoimprint pattern transfer and lithography. Using electron-beam evaporation, a silicon substrate
was layered with $5$ nm of Ti as adhesion and $200$ nm of Au to form the grating bulk material. A nanoimprint resist
bilayer was spun and soft-baked on top with $75$ nm of PMMA forming the underlayer and $160$ nm of upper layer resist
(AR-UVP), onto which a polymeric mold containing the grating features was aligned and stamped while curing the
imprinted resist (EVG aligner). After mold/sample separation, reactive ion etch (RIE) recipes were used to remove
residual top and bottom layer resists within the pattern trenches. The pattern was transfered from imrpint resist to a
metal mask by $30$ nm of Cr deposition and acetone liftoff. Using the Cr mask, an additional RIE recipe was used to
directly remove $30$ nm of Au within the exposed trenches, followed by wet etch removal of Cr. The pattern depth, duty
cycle, and periodicity were confirmed by AFM measurements (Nanonics).

For droplet generation in white light imaging setup we used $25.4$ mm diameter Axicon (Thorlabs AR coated UC fused silica) with deflection angle of $0.23^{o}$ and with reflectence less than $1$ \% in the range $350$ - $700$ nm.

To form a TLD film, silicone oil (Fisher Scientific) of refractive index $1.39$ was spun onto the grating by repeated intervals of spin coating at $10,000$ rpm. Baseline thickness of the prepared fluid film was measured by spectral reflectance to be on average $176$ nm by a separate optical profilometer (Filmetrics F20). A complete spin curve for $3-12$ each spin being $1.5$ min in duration, was obtained to prepare baseline average fluid thickness from $175$ to $700$ nm.

The two complementary imaging modalities we employed in this work, i.e. real-space imaging and k-space imaging employ a very similar optical design described by Fig.S4 in Supplementary Material. In both imaging methods the low power $\lambda = 785$ nm probing-beam and the higher power $\lambda = 488, 514$ nm heating-beam, were brought to the same optical path by means of a short pass filter (SPF). Upon reflection from the sample, the light was imaged to CMOS camera by utilizing a 50:50 beam splitter (BS). Common to both imaging techniques is microscope imaging system with long working distance 50X objective and internal tube lens. The real-space imaging is accomplished by placing the CMOS camera at the focal plane of the tube lens. For k-space imaging we employed additional lens in order to image the BFP (see Supplementary Material for more details).
The standard figure of merit for sensitivity definition is to take it as three times of sigma, where sigma is defined as the standard deviation of the signal. In our case it is natural to compute sigma as the standard deviation of the fluctuating signal as a function of time, with the heating beam switched-off and the probing beam is switched-on; direct computation yields sigma equal to $0.2$ nm and the resultant sensitivity is then $0.6$ nm.

\section*{Funding Information}

This work was supported in part by the National Science Foundation (NSF), the Office of
Naval Research (ONR), the Semiconductor Research Corporation (SRC), the Army Research Office (ARO), DARPA, Cymer Corporation,
and the San Diego Nanotechnology Infrastructure (SDNI) supported by the NSF
National Nanotechnology Coordinated Infrastructure (grant ECCS-1542148).

%\section*{Acknowledgments}

\section*{Supplemental Documents}

%See  supplemental documents for derivation details and an experimental movie file desribing thin liquid film thickness as a function of time. 

See Supplement Material attached below.

%\bigskip \noindent See \href{link}{Supplement 1} for supporting content.

% Bibliography
%\bibliography{sample}

% Full bibliography will be added automatically on a new page for Optics Letters submissions. This command is ignored for journal article submissions.
% Note that this extra page will not count against page length.
%\bibliographyfullrefs{sample}

%Manual citation list

%%%%%%%%%% Merge with supplemental materials %%%%%%%%%%
% \pagebreak
\widetext
\begin{center}
\newpage
\title{SI}
\textbf{Supplemental Material for \\
``Subnanometer imaging and controlled dynamical patterning of thermocapillary driven deformation of thin liquid films''}

\text{Shimon Rubin, Brandon Hong and Yeshaiahu Fainman}

\textit{Department of Electrical and Computer Engineering, University of California, San Diego, 9500 Gilman Dr., La Jolla, California 92023, USA}
\end{center}
%%%%%%%%%% Merge with supplemental materials %%%%%%%%%%

\setcounter{equation}{0}
\setcounter{figure}{0}
\setcounter{section}{0}
\setcounter{table}{0}
\setcounter{page}{1}

\renewcommand{\thesection}{S.\arabic{section}}
\renewcommand{\thesubsection}{\thesection.\arabic{subsection}}
\makeatletter 
\def\tagform@#1{\maketag@@@{(S\ignorespaces#1\unskip\@@italiccorr)}}
\makeatother
\makeatletter
\makeatletter \renewcommand{\fnum@figure}
{\figurename~S\thefigure}
\makeatother
\makeatletter \renewcommand{\fnum@table}
{\tablename~S\thetable}
\makeatother
%\makeatletter \renewcommand{\fnum@equation}
%{\equationname~S\theequation}
%\makeatother

\section{Momentum conservation diagrams}

\begin{figure}[h]
	\includegraphics[scale=0.19]{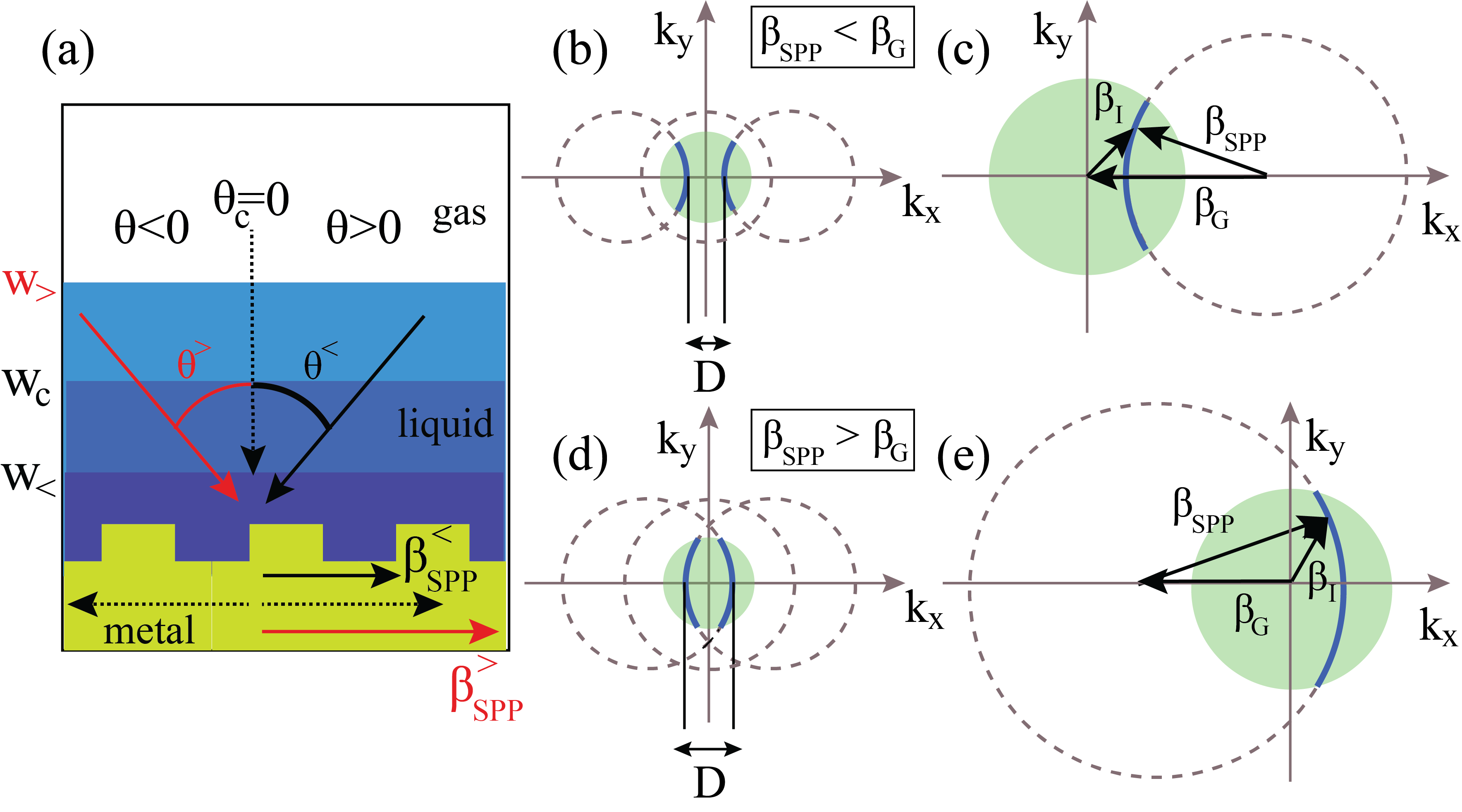}
    \caption{
        (a) Describes grating-coupling regimes into an SPP mode; thin film regime of thickness $w<w_{c}$, and thicker film regime of thickness $w>w_{c}$, where $w_{c}$ is the critical thickness which corresponds to the normal angle coupling. 
    (b,d) Present the allowed SPP dark arcs formed by an intersection of the set of incident light directions (green disks) with the shifted SPP momenta for the two regimes $\beta_{SPP}<\beta_{G}$ and $\beta_{SPP}>\beta_{G}$. (c,e) Present the momentum balance for the two regimes where the excited SPP mode carries in-plane momenta in the direction opposite (c) and parallel (e) relative to the in-plane component of the incident light momentum.}
    \label{KSpaceCircles}
\end{figure}

Fig. S\ref{KSpaceCircles} presents schematic description of the momentum conservation relation, given by Eq.3 in the main text.

The latter, describes coupling of an incident light into SPP mode of momentum $\beta_{SPP}$ which allows to deduce the corresponding resonant coupling angle $\theta$.
In particular, employing Eq.5 in the main text as well as the mutual orientation of the emerging two dark resonant arcs, allows to determine $\theta$ by experimental measurement of the distance $D$ between the arcs.
For films thinner than the critical thickness $w<w_{c}$, the emerging resonant dark arcs, point towards each other (Fig. S\ref{KSpaceCircles}(b)), whereas for thicker films, i.e. for $w>w_{c}$, the dark arcs point in the opposite directions. Furthermore, while in the $w<w_{c}$ regime the in-plane component of an incident free-space mode is anti-parallel to $\vec{\beta}_{SPP}$ ($\theta>0$), the $w>w_{c}$ regime is characterized by $\theta<0$ and SPP mode momentum which is parallel to the in plane component of an incident plane-wave momentum. Importantly, some cases, such as the $\lambda=1064$ nm probing-beam case described in Fig. 2(g) in the main text, do not admit critical height $w_{c}$. 
Consequently, the relative orientation of the emerging dark arcs in such cases is oriented towards each other for all thickness values. 
For sufficiently thick film, above a WG threshold condition, the dielectric film and the metal grating support also resonant coupling into WG modes.
The corresponding resonant angle diagrams would then admit an additional pair of dark arcs (one pair for each higher-order WG mode) on top of the resonant dark arcs associated with the SPP mode. Small thickness perturbations of such thick films would lead to changes of the WG dark arcs whereas the SPP resonant curves would not change as could be also anticipated from the low sensitivity regions of SPP curves at high film thicknesses, presented in Fig. 2 in the main text.

%It is worth mentioning that despite the fact that SPP and WG modes both propagate parallel to the metal surface and both are coupled from the free-space by diffraction from the metal grating, measuring thickness changes due to changes of WG modes momentum should not be considered as an extension of the SPRM but rather as an independent method. The main reason for such distinction is because the propagation of the WG modes is set by its interaction with dielectric, and not with a metal as in the case of the SPP mode.

\section{Underlying main assumptions of Eq.5}

The free surface of a Newtonian liquid film of viscosity $\mu$ and stress tensor $\tau_{ij}$ satisfies the following stress balance equation \cite{oron1997longSM}
\begin{equation}
	\tau_{ij} n_{j}  = \sigma n_{i} \vec{\nabla} \cdot \hat{n} - \vec{\nabla}_{\parallel} \sigma; \quad i,j=x,y,z
\label{MatchingConditions}    
\end{equation}
where, $\sigma$ is the surface tension,  $\vec{\nabla}_{\parallel}$ stands for a gradient with respect to the in-plane coordinates ($y,z$) 
and $\vec{\nabla} \cdot \hat{n}$ is the 3D divergence of the unit vector normal to the film surface. 
%Assuming that the surface tension depends linearly on the temperature via \cite{levich1962physicochemical}
%\begin{equation}
%	\sigma(T)=\sigma_{0}-\sigma_{T} \Delta T; \quad \Delta T \equiv T-T_{0},
%\label{SurfaceTensionGrad}
%\end{equation}
%where $\sigma_{T}$ is the so-called Marangoni constant 
%known to exhibit nearly temperature independent values over a large range of temperature for many materials \cite{adamson1990physical}.
Assuming the surface tension depends linearly on the temperature via $\sigma(T)=\sigma_{0}-\sigma_{T} \Delta T$, applying lubrication approximation
%\cite{howison2005practical}, 
for the Navier-Stokes equation, 
%which allows to drop the in-plane derivatives relative to the normal derivative and drop the inertial terms, 
taking the thin film limit
%\cite[ch.\ 20]{howison2005practical} 
of the matching conditions on the fluid-gas interface \cite{oron1997longSM}, 
and assuming quasi-static temperature field distribution, 
%which expected to hold to a great degree of accuracy due to short heat diffusion time relative to the thin film deformation time, 
yields \cite{rubin2018nonlocalSM}
\begin{equation}
	\dfrac{\partial \eta}{\partial t} + D_{\sigma} \nabla^{4}_{\parallel}  \eta = - \dfrac{\sigma_{T} w_{0}^{2}}{2 \mu}  \dfrac{\Delta T}{I_{0} d^{2}}  \chi  I.
\label{ThinFilmQuasiStat}    
\end{equation}
%The latter includes the effects of surface tension and thermocapillarity, but doesn't include effects of gravity which are negligible on a microscale and expected to emerge on a much larger scales.
The minus (positive) sign of the source term in Eq.(\ref{ThinFilmQuasiStat}) indicates decrease (increase) of local thickness for positive (negative) Marangoni constant.
Scaling to dimensionless variables via $t \rightarrow \tau_{l} t$ where $\tau_{l}=d^{4}/D_{\sigma}$, $\vec{r}_{\parallel} \rightarrow (\tau_{th} \cdot D_{\sigma}^{1/4}) \vec{r}_{\parallel}$, $\eta \rightarrow w_{0} \eta$ and $I \rightarrow I/I_{0}$ yields
\begin{equation}
	\dfrac{\partial \eta}{\partial t} + \nabla^{4}_{\parallel}  \eta =  - \text{Ma} \cdot  \chi \cdot \frac{\tau_{l}}{\tau_{th}}  I/2,
\label{ThinFilmQuasiStatNonDim}    
\end{equation}
where $\text{Ma} \equiv \sigma_{T} \Delta T w_{0}/(\mu D_{th}^{m})$ is the dimensionless Marangoni number. 

%%%%%%%%%%%% Estimate For Maximal Temperature %%%%
The heat diffusion equation
that governs the temperature field, $T^{m}$, on  a metal substrate with heat diffusivity $D_{th}^{m}$ is
\begin{equation}
	\dfrac{\partial T^{m}}{\partial t} - D_{th}^{m} \nabla^{2}_{\parallel} T^{m} =  \dfrac{\Delta T}{I_{0} \tau_{th}}  \chi  I; \quad \chi \equiv \dfrac{\alpha_{th}^{m}  d^{2} I_{0}}{k_{th}^{m} \Delta T},
\label{HeatEquationDiffusion2c}.    
\end{equation}
%of the free surface is identical to the temperature field in the metal, $T^{m}$, which is governed by the following 2D equation
The latter implies that the maximal temperature arise, $\Delta T^{max}$, due to a circular laser beam of intensity $I$ and waist $\ell_{0}$, is subject to
% \cite{bauerle2013chemical}
\begin{equation}
	\Delta T^{max} = \dfrac{\alpha I \ell_{0}}{k_{th}^{m}},
\end{equation}
where $\alpha$ is the optical absorption coefficeint.
For a low power beam of intensity $I=19.98 \cdot 10^{7}$ Wm$^{-2}$ and heat conductivity of gold, which is around $k_{th}^{m}=300$ Wm$^{-1}$K$^{-1}$, reflectance at $\lambda=514$ nm and $\ell_{0}= 0.8$ $\mu$m, 
%\cite{bauerle2013chemical},
the corresponding $\Delta T^{max} = 0.25$ K.
%Here, we assumed that the local temperature of TLD film is identical to the temperature of the metal substrate.

\section{Derivation of Eq.6}

The governing equation for the axisymmetric flow of a TLD Newtonian film under the effects of surface tension and centrifugal force due to rotation of the substrate with axisymmetric roughness, and furthermore under lubrication approximation is given by \cite{stillwagon1990levelingSM}
\begin{equation}
	r \dfrac{\partial h}{\partial t} = -\dfrac{1}{3 \mu} \dfrac{\partial}{\partial r} \left(\sigma  h^{3} r \dfrac{\partial^{3} (h+s)}{\partial r^{3}}  + h^{3} \rho \omega^{2} r^{2} \right).
\end{equation}
Here, $r$ is the radial coordinate, $\rho$ is the fluid density, $\omega$ is the angular velocity, $h$ is the local TLD film thickness relative to the topographic feature and $s$ is the local substrate height, so $h+s$ is the local height of the fluid-gas interface relative to the substrate. Switching to dimensionless variables via
\begin{equation}
\begin{split}
	&R \equiv \dfrac{r}{w-q}, X \equiv \dfrac{r-r_{0}}{w-q}, H \equiv \dfrac{h}{w-q}, S \equiv \dfrac{s}{w-q}, 
\\
	&T \equiv \dfrac{t}{t_{c}}, t_{c} \equiv \dfrac{\Lambda \mu}{2 r_{0} \rho \omega^{2} (w-q)^{2}}
\end{split}
\end{equation}
where $w-q$ represents fluid thickness above the grating ridge and $t_{c}$ is the time scale based on the centrifugal force, and
furthremore dropping the time derivative yields the following time independent equation for dimensionless thin film thickness \cite{stillwagon1990levelingSM}
\begin{equation}
	\left( \dfrac{\partial^{3} H}{\partial X^{3}} + \dfrac{\partial^{3} S}{\partial X^{3}} \right) H^{3} + \Omega^{2} H^{3} = \Omega^{2}; \quad \Omega^{2} \equiv \dfrac{\rho \omega^{2} (\Lambda/2)^{3}  r_{0} }{\sigma (w-q)},
\label{NodDimRot}	
\end{equation}
where $\Omega^{2}$ represents a dimensionless ratio of centrifugal to capillary forces. 
Here, dropping the time derivative is justified because the characteristic time for the film thinning process, $t_{s} \equiv 2r_{0} t_{c}/\Lambda$, is much larger than that for the rotational flow $t_{c}$, and therefore we can expect the film profile over the grating to adjust to changes in the overall film thickness at a rate much faster than the overall film thickness itself changes \cite{stillwagon1990levelingSM}. In fact, in our system  $\Lambda=6 \cdot 10^{-7}$ m, $r_{0}=10^{-3}$ m and $2r_{0}/\Lambda \gg 1$. 
Treating to $S$ as a piece-wise constant function, and linearizing Eq.(\ref{NodDimRot}) above the grating ridge and trench via $H=1+ \eta$ and $H=1+q/(w-q)+\eta$, respectively, we can omit the derivative of $S$ leading to \cite{stillwagon1990levelingSM}
\begin{subequations}
\begin{align}
	&\dfrac{\partial^{3} \eta}{\partial x^{3}} +3  \Omega^{2} \eta = 0,
\\
	&\dfrac{\partial^{3} \eta}{\partial x^{3}} + \Omega^{2} \Big[ 1 - \left( 1 + \dfrac{q}{w-q} \right)^{-3} + 3 \left(1 + \dfrac{q}{w-q} \right)^{-4} \eta  \Big] = 0.
\end{align}
\label{NoddomOmega}
\end{subequations} 

\section{Details of numerical simulation}

\textbf{Geometry}: The scheme of the simulation domain is presented at Fig. 1 in the main text and it resides in the region $-0.3$ $\mu$m $\leq x \leq 0.3$ $\mu$m and $-3$ $\mu$m $\leq y \leq$ $1.5$ $\mu$m. The metal substrate resides in the region between $-3$ $\mu$m $\leq y \leq 0$ $\mu$m and it hosts another $30$ nm depth metal grating stripe in the region $-0.15$ $\mu$m $\leq x \leq 0.15$ $\mu$m (i.e. half-duty cycle) and $-0.33$ $\mu$m $\leq y \leq -0.3$ $\mu$m.

\textbf{Computational domain}: The computational grid accommodates $67$ and $279$ elements along the $x$- and $y$-axis, respectively. While the spacing between numerical grid points is uniform along the $x$-axis, the grid is non-uniform along the $y$- axis. The latter admits a finer mesh near the metal surface, accommodating $35$ elements in the region near the metal surface, $-0.33 \text{ }\mu \text{m} \leq y \leq -0.3$ $\mu$ m. The implemented boundary conditions are Bloch along $x$-axis and PML along $y$ and $z$ axes. PML settings: stretched coordinate PML; min layer: 12; max layers: 64; kappa=3; sigma=1.5; polynomial:3; alpha=0; alpha polynomial:1. 

\textbf{Dielectric properties:} 
For incident plane wave of wavelength $\lambda=785$ nm and $\lambda = 1064$ nm, the corresponding dielectric constant of gold is $\epsilon_{m} = -22.855+ 1.4245 i$ and $\epsilon_{m} = -48.450 + 3.6006 i$, respectively  \cite{johnson1972opticalSM}. 
The refractive index of the thin film is $1.39$, which corresponds to refractive index of silicone oil employed in our experiments.

\textbf{Source and monitor:} The source is a plane wave injected at $y=-1.7$ $\mu$m at different angles, typically between $0$ and $20$ relative to the $y$ axis. The monitor collects the reflected wave at $y=-2.2$ $\mu$m.

\section{Numerical results of an angular band-gap and the effect of thin film's curvature on the resonant coupling angle}

Increasing the dielectric thickness $w$ on top of the metal grating from zero thickness (i.e. no dielectric) to higher values, leads to a monotonic increase of the coupled SPP wavevector, and therefore results in an increasingly smaller resonant coupling angle. In some cases where the coupling angle at $w=0$ is sufficiently low, the dielectric angular band-gap opens around the normal incidence angle at critical thickness $w_{c}$, which results in a lower coupling efficiency into SPP mode. Fig. S\ref{AngBandGap} below, describes coupling of a TM polarized $785$ nm plane-wave into SPP mode on gold grating (grating period $600$ nm) with an angular bandgap around critical thickness $w_{c}=133$ nm (Fig.S2(b)). 

\begin{figure}[t!]
	\includegraphics[scale=0.33]{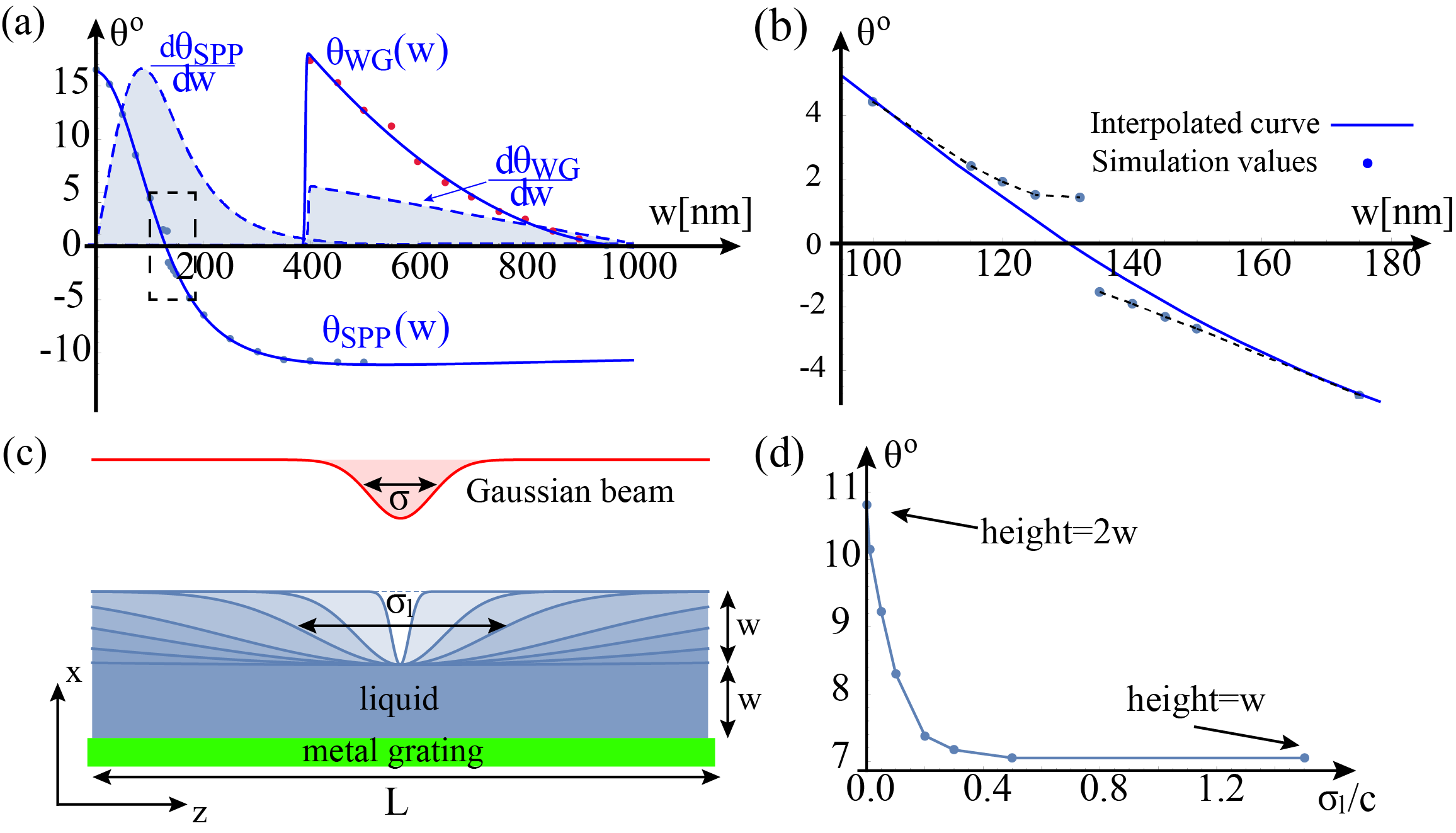}
    \caption{Numerical results presenting: (a) coupling angle curves to SPP and WG modes for probing-beam of wavelength $\lambda = 785$ nm and (b) angular band-gap formed around the normal incidence angle at critical thickness around $w=133$ nm. 
    (c) Surface topographies described by Eq.\ref{GaussianDimple} for the values $\sigma_{l}/c=0,0.01,0.05,0.1,0.2,0.3,0.5,1.5$, $L=21$ $\mu$ m, $w=225$ $\mu$ m, distance of source from the metal $850$ nm, $\sigma=3$ $\mu$ m, and $c=42 \cdot 10^{-6}$ dimensionless constant. 
    (d) The resonant coupling angle of the Gaussian probing beam waist $\sigma$, as a function of the indentation width. For sufficiently wide (narrow) indentations the coupling angle tends to the resonant angle of flat film of thickness $w$ ($2w$).}
    \label{AngBandGap}
\end{figure}

We would like to probe the effect of local non-homogeneity in TLD film thickness, on the value of the corresponding resonant coupling angle to the SPP mode. To this end, we construct a numerical simulation domain of width $L$, schematically presented in Fig. S2c, and employ a commercial-grade simulator based on the finite-difference time-domain method (Lumerical, FDTD). The shape of the non-homogeneity is taken as an indentation of a Gaussian form, with a width parametrized by $\sigma_{l}$ via
\begin{equation}
	x(z)=w[2 - \exp(-z^{2}/\sigma_{l}^{2})].
\label{GaussianDimple}
\end{equation}
% which takes values both smaller and larger than the width at half maximum of the the Gaussian beam, $\sigma$.
Fig. S\ref{AngBandGap}d presents numerical results of the corresponding resonant angle as a function of indentation width for a fixed  Gaussian source beam, $\exp(-x^{2}/\sigma^{2})$, of width at half maximum $\sigma$.
In the limit of increasingly wider indentations, described by $\sigma_{l}>\sigma$, the coupling angle tends to the resonant angle that corresponds 
to a flat dielectric of thickness $w$, whereas in the opposite limit of narrower indentations ($\sigma_{l}<\sigma$), the coupling angle tends to the resonant angle that corresponds to a flat dielectric of thickness $2w$. In particular, we learn that for increasingly smaller $\sigma_{l}$ the coupling angle experiences a significant shift only at sufficiently sharp indentations that introduce a mean slope around the value $2w/L \simeq 0.02$. 

\section{Numerical simulation details in Fig.4}

Numerical integration of Eq.(\ref{ThinFilmQuasiStatNonDim}) is accomplished by utilizing commercial grade numerical solver (Mathematica, v. 11) and implementing its built-in Explicit Runge Kutta method. The size of the simulation domain in dimensionless units along the $x$ and $y$ directions is $30$, 
and the deformation is subject to a vanishing boundary conditions on the lines $y,z=\pm 15$. The spacing of the numerical grid is 0.3 and the number of effective digits of precision is three. The dimensionless number $\text{Ma} \cdot \chi \cdot \tau_{l}/(2\tau_{th})$ is taken as unity. 
The intensity, $I$, is taken as $I(r)=\exp(-(r-r_{0})^{2})$, where $r_{0}=0$ corresponds to maximal temperature and thinning at the origin, whereas 
% (t=5)    
$r_{0}=5$ corresponds to maximal temperature and thinning on a circle of radius $r_{0}$, which leads to a drop-like structure. 

\section{Comparison of sensitivities to thickness changes due to Fresnel reflection and coupling into SPP}

Consider the value of the ratio $s$, defined as $s \equiv \Delta R/\Delta w$ where $\Delta R$ is the change of reflectivity due to the corresponding change of TLD film thickness and $\Delta w$ is the change of thickness whereas. For Fabry-Perrot (FP) and SPP modes, the maximal value of the ratio $s$ is given by $s_{max}^{FP}=\Delta R^{FP}_{max}/(\lambda/4n_{l})$ and $s_{max}^{SPP}=\Delta R^{SPP}_{max}/\Delta w (\Delta \theta_{1/2})$, respectively. Here, $\lambda/4n_{l}$ is the change of the dielectric thickness that corresponds to a maximal shift of reflectivity given by  $\Delta R^{FP}_{max}=4 r^{2}/(1+r)^{2}$ where $r=\vert (n_{g}-n_{l})/(n_{g}+n_{l}) \vert$ is the Fresnel normal reflection coefficient; $\Delta w (\Delta \theta_{1/2})$ is the change of thickness which corresponds to twice the width of the angular half-width depth, $\Delta \theta_{1/2}$, of the corresponding resonance angle -thickness curve. Assuming $\Delta R^{SPP}_{max}=1$, and furthermore utilizing the values $n_{g}=1$, $n_{l}=1.39$, $\lambda=785$ nm, $\Delta \theta_{1/2}=4^{o}$ and the corresponding $\Delta w (\Delta \theta_{1/2})=15$ nm, yields enhancement of the proposed method by $s_{max}^{SPP}/s_{max}^{FP} \simeq 15$.

\section{Shape of thin liquid film under spinning on a substrate with periodic topography}

Following \cite{wu1999completeSM}, the corresponding solution of the governing equation (Eq.(14) in the main text) of TLD film under capillary and centrifugal forces in the quasi-static limit is given by the linear superposition of the functions $\varphi_{1,2,3,4}(X)$
\begin{equation}
\begin{split}
	H_{t} &= 1 + \dfrac{q/(w-q)}{1+\Omega^{2}} + a_{t} \varphi_{1}(X) + b_{t} \varphi_{2}(X) + c_{t} \varphi_{3}(X) + \varphi_{4}(X)
\\
	H_{r} &= 1 + a_{r} \varphi_{1}(X) + b_{r} \varphi_{2}(X) + c_{r} \varphi_{3}(X)
\end{split}
\end{equation}
Here, $t,r$ denote regions above the grating trench and ridge, respectively, and the corresponding functions are given by
\begin{equation}
\begin{split}
	\varphi_{1}(X) &= e^{-\lambda_{r,t}X}
\\
	\varphi_{2}(X) &=  e^{\lambda_{r,t}X/2} \cos \left( \sqrt{3} \lambda_{r,t} X/2 \right)
\\
	\varphi_{3}(X) &= e^{\lambda_{r,t}X/2} \sin \left( \sqrt{3} \lambda_{r,t} X/2 \right)
\\
	\varphi_{4}(X) &=	\dfrac{1}{3} \Big[ \left( 1 + \dfrac{q/(w-q)}{1+\Omega^{2}} \right) - \left( 1 + \dfrac{q/(w-q)}{1+\Omega^{2}} \right)^{4} \Big]
\end{split}	
\end{equation}
where $\lambda_{r,t}$ take the following values in each one of the domains
\begin{equation}
	\lambda_{r}= (3 \Omega^{2})^{1/3} ; \quad \lambda_{t} =  \left( 3 \Omega^{2} \left( \dfrac{q/(w-q)}{1+\Omega^{2}} \right)^{-4} \right)^{1/3}.
\end{equation}
The corresponding coefficients $a_{t,r}$, $b_{t,r}$, $c_{t,r}$ are then determined by the continuity of the fluid-air height as well as its first and second derivatives at points $X=-1/2$ and $X=(1+\Lambda)/2$.

Inserting the value $\Omega^{2} = 2.06 \cdot 10^{-4} \ll 1$ derived below Eq.(14) in the main text, we find that under surface tension and centrifugal forces the following values of mean thickness 
%$w=$$36, 60, 90, 150, 270$ 
$w=$$36, 60, 90, 150$ 
nm lead to the following values of 
%$H=$ $5.9 \cdot 10^{-7}$, $3.6 \cdot 10^{-5}$, $1.2 \cdot 10^{-5}$, $3.9 \cdot 10^{-6}$, $8.9 \cdot 10^{-7}$ 
$H=$ $5.9 \cdot 10^{-7}$, $3.6 \cdot 10^{-5}$, $1.2 \cdot 10^{-5}$, $3.9 \cdot 10^{-6}$ 
in units of $w-q$, 
i.e. much smaller than $1$ nm.

\section{Numerical results of the effect of thin film's undulation on the value of the resonant coupling angle}

\begin{figure}[th!]
	\includegraphics[scale=0.23]{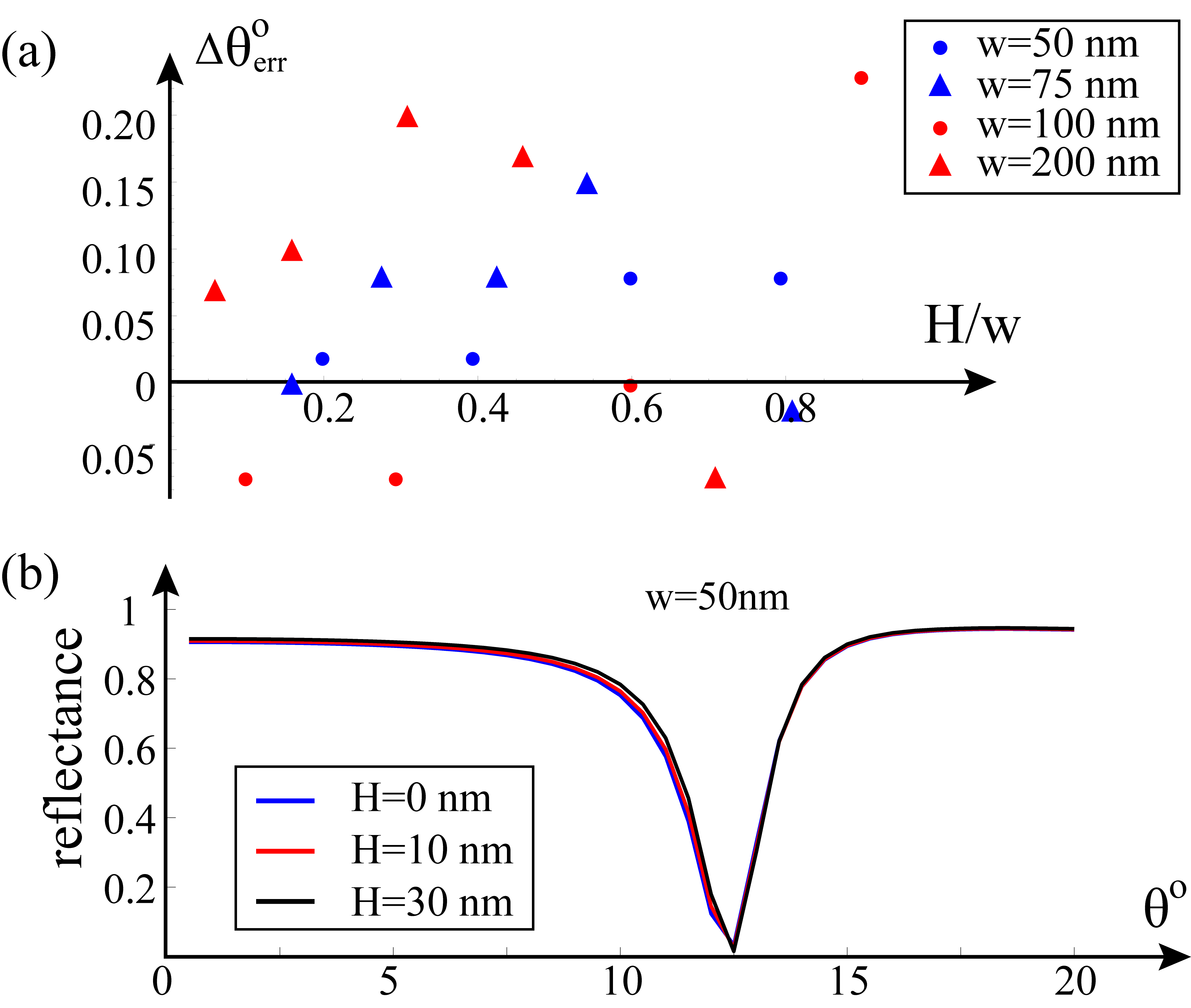}
    \caption{Numerical simulation results presenting the effect of the periodic undulation amplitude on the resonant coupling angle. (a) Presents the value of the resonant angle as a function of the dimensionless undulation peak to peak height, $H/w$, and (b) presents the reflectance curve for TLD films of thickness $50$ nm with different peak to peak undulation with values  $H=0, 10, 30$ nm, indicating that periodic and symmetric undulation does not introduce a significant shift to the resonant coupling angle.}
    \label{1064Sensitivity}
\end{figure}

Let us determine the effect of thin film undulation derived above, on the value of the resonant coupling angle. To this end we perform numerical simulations of an incoming plane wave on a grating of period $\Lambda$ described in Fig. 4(a) in the main text. The corresponding shift in the resonant angle is described in Fig. S\ref{1064Sensitivity}a. Remarkably, the corresponding angle shift is a very weak function of $H/w$, even for values of $H$ comparable to 
$w$. Fig. S\ref{1064Sensitivity}b furthermore indicates that the reflectance curve is practically unaffected by the periodic symmetric undulation. Periodic undulations which are not symmetric relative to the grating are expected to modify the resonant coupling angle for sufficiently large amplitude, and these are not considered in this work. 

\section{Experimental setup, fabrication details, and sample preparation}

\begin{figure}[th!]
	\includegraphics[scale=0.29]{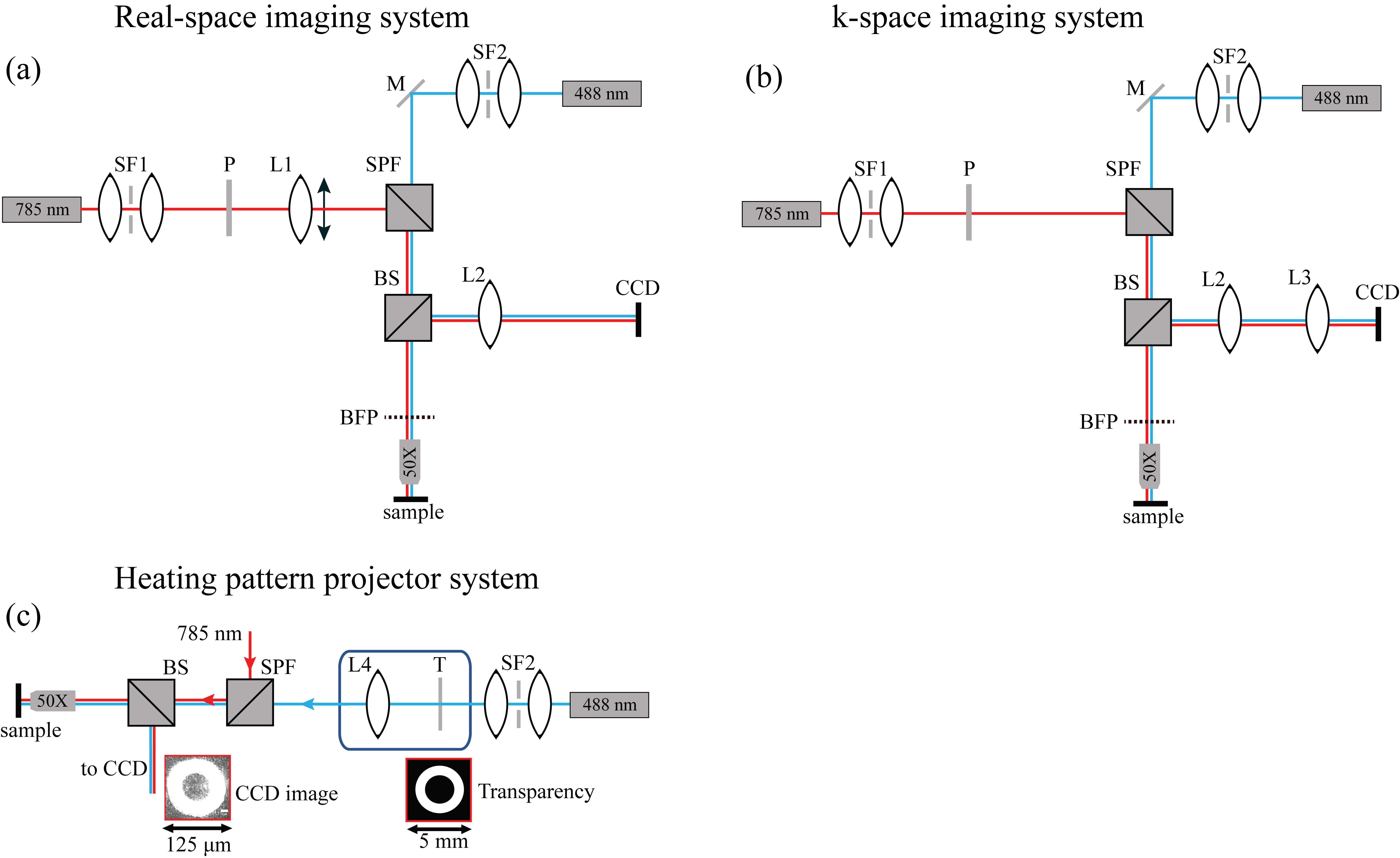}
    \caption{Experimental setup for: (a) real-space and (b) k-space imaging systems, and (c) the pattern projection setup to induce droplet directly from TLD film. SF1 and SF2 stand for spatial filters (SFs) comprised from a pair of lenses and a pinhole.  Real-space imaging is achieved by placing the camera in the back focal plane (BFP) of the tube lens, and by placing an additional lens L1 in the illumination path of the probing beam in order to deliver a plane wave to the sample; L1 is mounted on a translation stage which can change its position in a direction perpendicular to the optical axis. k-space imaging is achieved by inserting lens L3 after lens L2 in order to complete a 4f imaging system to image the BFP of the microscope objective, and by removing lens L1 in order to focus the probing beam onto the sample. Other components: SPF - short pass filter, P - polarizer, L2 - tube lens, BS - 50:50 beam splitter. The heating pattern projector unit described in (c) is comprised of a transparency mask and additional 15 inch lens L4 placed before SPF.}
    \label{ExpScheme}
\end{figure}

Fig.S\ref{ExpScheme} presents the schematic description of the optical setup we employed in our experiments.
In particular, Fig.S\ref{ExpScheme}a and Fig.S\ref{ExpScheme}b describe the experimental setups we used for real-space and $k$-space measurements, respectively, whereas Fig. S\ref{ExpScheme}c presents the pattern projection setup. The low power near infra-red (NIR) laser diode ($\lambda = 785$ nm) is used as a probing-beam, whereas the higher power argon laser ($\lambda = 488,$ $514$) nm is used as a heating-beam; these two beams were brought to the same optical path by means of a short pass filter (SPF). Upon reflection from the sample, the light was imaged to CMOS camera by utilizing a 50:50 beam splitter (BS). 
 Real-space imaging is achieved by placing the camera in the the back focal plane (BFP) of the tube lens, and by placing an additional lens L1 in the illumination path of the probing beam in order to deliver a plane wave to the sample; L1 is mounted on a translation stage which can change its position in a direction perpendicular to the optical axis. $k$-space imaging is achieved by inserting lens L3 after lens L2 in order to complete a 4f imaging system to image the BFP of the microscope objective, and by removing lens L1 in order to focus the probing beam onto the sample.
% Real-space imaging is achieved by placing additional lens L3 in order to image the back focal plane (BFP) to the CMOS camera (i.e. L2 and L3 form a 4f imaging block), and by placing an additional lens L1 in order to deliver a plane wave to the objective; L1 is mounted on a translation stage which can change its position in a direction perpendicular to the optical axis (marked by an arrow next to L1). $k-$ space imaging is achieved by removing L3 and by adjusting the camera location along the optical path in order to image sample's surface, and by removing L1 to focus the probing beam on the BFP. 
The heating pattern projector unit described in Fig. S\ref{ExpScheme}c, is comprised of a transparency mask and an additional 15 inch lens L4 placed before SPF. For droplet generation in white light imaging setup we used $25.4$ mm diameter Axicon (Thorlabs AR coated UC fused silica) with deflection angle of $0.23^{o}$ and with reflectence less than $1$ \% in the range $350$ - $700$ nm, and heating beam which employs $400$ mW laser power source of wavelength $\lambda=532$ nm.

For the metal substrate which supports SPP excitations and hosts TLD film we employed a  $50$\% duty cycle gold grating of $600$ nm period and with $30$ nm depth grooves which were fabricated by nanoimprint pattern transfer and lithography, as described below. Using electron-beam evaporation, a silicon substrate was layered with $5$ nm of Ti as adhesion and $200$ nm of Au to form the grating bulk material. A nanoimprint resist
bilayer was spun and soft-baked on top with $75$ nm of PMMA forming the underlayer and $160$ nm of upper layer resist
(AR-UVP), onto which a polymeric mold containing the grating features was aligned and stamped while curing the
imprinted resist (EVG aligner). After mold/sample separation, reactive ion etch (RIE) recipes were used to remove
residual top and bottom layer resists within the pattern trenches. The pattern was transfered from imrpint resist to a
metal mask by $30$ nm of Cr deposition and acetone liftoff. Using the Cr mask, an additional RIE recipe was used to
directly remove $30$ nm of Au within the exposed trenches, followed by wet etch removal of Cr. The pattern depth, duty
cycle, and periodicity were confirmed by AFM measurements (Nanonics).

To form a TLD film, silicone oil (Fisher Scientific) of refractive index $1.39$ was spun onto the grating by repeated intervals of spin coating at $10,000$ rpm. Baseline thickness of the prepared fluid film was measured by spectral reflectance to be on average $176$ nm by a separate optical profilometer (Filmetrics F20). A complete spin curve for $3-12$ each spin being $1.5$ min in duration, was obtained to prepare baseline average fluid thickness from $175$ to $700$ nm.

\section{Movie demonstrating $k$-space thickness measurement of TLD film}

Experimental $k$-space imaging results, demonstrating dynamical nanometric thickness change of thin silicone oil film as a function of time due to optically-driven TC effect and the following healing dynamics after the illumination source is switched off. Top, presents the resonant angles along the central line of the $k$-space diagram, whereas bottom present the corresponding thickness at the illuminated small region, nearly of minimal size allowed by diffraction.

%\begin{condenseditemize}
%\item[] Algorithm S1
%\item[] Equation (S1)
%\item[] Figure S1
%\item[] Media S1
%\item[] Table S1
%\end{condenseditemize}

%\section*{Media}

%The supplemental document may contain linked objects such as video, 2D, 3D, and machine-readable data files. Please see the \href{http://www.opticsinfobase.org/submit/style/supplementary-materials-optica.cfm}{Author Guidelines for Supplementary Materials} for more information. Such files should be cited in the supplementary document as in the primary document but using the naming convention described above.

%\section*{References} 

% Bibliography
%\bibliography{sample}

%Manual citation list

\end{document}